\documentclass[11pt]{article}
\usepackage{geometry}                
\usepackage{setspace}
\usepackage[parfill]{parskip}    
\usepackage{newtxtext}

\usepackage{url}
\usepackage{hyperref}
\usepackage{amsmath}
\usepackage{amssymb}
\usepackage{amsthm}
\usepackage{bm}

\usepackage{times}
\usepackage{epsfig}
\usepackage{graphicx}

\usepackage{booktabs}
\usepackage{multirow}
\usepackage{caption}
\usepackage{subcaption}
\usepackage[noend]{algpseudocode}
\usepackage{algorithm}

\title{Advanced Multi-Microscopic Views Cell Semi-supervised Segmentation}

\author{
Fang~Hu\thanks{F. Hu and X. Sun contributed equally to this work.}~\thanks{F. Hu, X. Sun, F. Xiao, Z. Wang, and X. Fan are with Moore Threads.}
\and Xuexue~Sun\footnotemark[1]~\footnotemark[2]~\thanks{Correspondence should be addressed to X. Sun \href{mailto:xuexue.sun@mthreads.com}{xuexue.sun@mthreads.com}.}
\and Ke~Qing\thanks{K. Qing is with University of Science and Technology of China.}
\and Fenxi~Xiao\footnotemark[2]
\and Zhi~Wang\footnotemark[2]
\and Xiaolu~Fan\footnotemark[2]
}

\date{}

\begin{document}

\maketitle

\begin{abstract}
Although deep learning (DL) shows powerful potential in cell segmentation tasks, it suffers from poor generalization as DL-based methods originally simplified cell segmentation in detecting cell membrane boundary, lacking prominent cellular structures to position overall differentiating. 
  Moreover, the scarcity of annotated cell images limits the performance of DL models. Segmentation limitations of a single category of cell make massive practice difficult, much less, with varied modalities.
   In this paper, we introduce a novel semi-supervised cell segmentation method called \textbf{M}ulti-\textbf{M}icroscopic-view \textbf{C}ell semi-supervised \textbf{S}egmentation (MMCS), which can train cell segmentation models utilizing less labeled multi-posture cell images with different microscopy well. 
  Technically, MMCS consists of \textbf{Nucleus-assisted global recognition}, \textbf{Self-adaptive diameter filter}, and \textbf{Temporal-ensembling models}. Nucleus-assisted global recognition adds additional cell nucleus channel to improve the global distinguishing performance of fuzzy cell membrane boundaries even when cells aggregate. Besides, self-adapted cell diameter filter can help separate multi-resolution cells with different morphology properly. It further leverages the temporal-ensembling models to improve the semi-supervised training process, achieving effective training with less labeled data. Additionally, optimizing the weight of unlabeled loss contributed to total loss also improve the model performance.
  Evaluated on the Tuning Set of NeurIPS 2022 Cell Segmentation Challenge (NeurIPS CellSeg), MMCS achieves an F1-score of 0.8239 and the running time for all cases is within the time tolerance. 
\end{abstract}

\section{Introduction}
\label{sec:intro}
Cell segmentation is a critical step in a variety of downstream microscopy image-based biology and biomedical research such as cell grouping \cite{gencctav2012unsupervised,todman1997cell,chankong2014automatic}, tumor immune micro-environment quantization \cite{cheng2018identification,turkki2016antibody,finotello2019molecular}, and disease diagnosis \cite{conway2020kidney,alam2017alzheimer,kong2016robust}.
While traditional cell segmentation approaches \cite{wahlby2004combining,sadeghian2009framework,meijering2012cell,piorkowski2016statistical,anoraganingrum1999cell} usually require careful strategy designs and hyperparameters tuning but could hardly achieve strong segmentation performance when microscopic cell images are in low contrast, recent advances have demonstrated the powerful potential of applying deep learning technique into cell segmentation.
U-Net, a convolutional neural network that is originally designed for biomedical image segmentation \cite{ronneberger2015u,zhou2018unet++,falk2019u},  has been adapted to help promote high-quality cell segmentation \cite{schmidt2018cell,panigrahi2021misic,cutler2022omnipose}.
Meanwhile, inspired by the success of mask R-CNN framework \cite{he2017mask} in object segmentation, some works further leverage it as a backbone method for cell segmentation task \cite{johnson2018adapting,johnson2019automatic,lin2021effective}.

Training a high-performance deep learning model would usually need feeding it with plenty of high-quality annotated training data. However, in biomedical image processing, it is very challenging to obtain a sufficient amount of annotated data for cell segmentation due to the time and expertise required for annotation in the fields of biology and biomedicine. \cite{willemink2020preparing,rubin2008medical,philbrick2019ril}.
On the other hand, in the case of cell segmentation, optimizing the deep learning model solely for the detection of cell membrane boundaries may oversimplify the segmentation process and fail to utilize other prominent cellular structures that could assist in differentiating the cells \cite{panigrahi2021misic,johnson2018adapting,johnson2019automatic}.
Cell images usually consist of various categories of cells and multiple microscopic views, which are difficult to be effectively learned, therefore leading to poor generalization ability of deep learning models in cell segmentation tasks.

To tackle the challenge of lacking high-quality cell images and promote the performance of cell segmentation, in this paper, we propose  \textbf{M}ulti-\textbf{M}icroscopic-view \textbf{C}ell semi-supervised \textbf{S}egmentation, {\it i.e.} \textbf{MMCS}.
MMCS is a novel cell semi-supervised segmentation approach that can effectively learn cell segmentation models from less annotated multi-posture cell images of different microscopic views.
Technically, it consists of three key procedures:
(1) \textbf{Nucleus-assisted global recognition}. This procedure will adopt additional nucleus features into deep model training, which could facilitate the model to learn more knowledge from the overall cell image data. 
(2)  \textbf{Self-adaptive diameter filter}. This procedure will recognize the properties of different cell morphology saved to the training model and further guild cell images to segment, which enable a deep model to learn knowledge with strong diversity and thus help improve the generalization ability of the learned model.
(3)  \textbf{Temporal-ensembling models}. This procedure aims to enhance the precision and dependability of the pseudo labels of unlabeled data during the semi-supervised training process, facilitating efficient training with a reduced amount of labeled data.

\begin{figure*}[h]
  \centering
  \includegraphics[width=\linewidth]{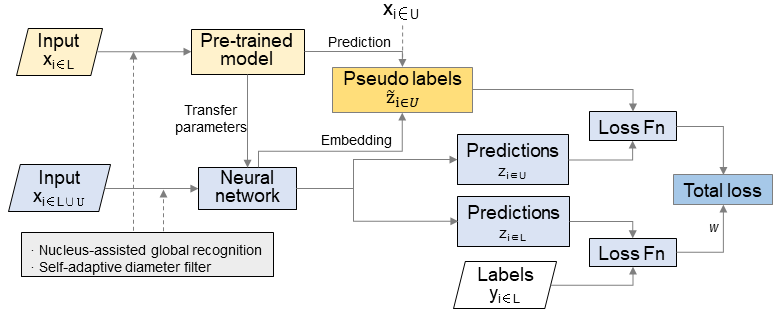}
  \caption{Overview of MMCS. Firstly, a pre-trained model is trained on all available labeled data to initialize pseudo labels for unlabeled data. And then, all the labeled and unlabeled data are fed into the network with parameters transferred from the pre-trained model. Features trained reweight the labeled and unlabeled loss and feedback the weight to adjust the knowledge importance. Along the learning process, the pseudo labels will be updated with the temporal-ensemble training model.}
  \label{fig:mmcs-workflow}
\end{figure*}

\renewcommand{\thefootnote}{\arabic{footnote}}
A simplified working pipeline of MMCS is summarized as Fig.\ref{fig:mmcs-workflow}.
Finally, we empirically evaluate the proposed MMCS on NeurIPS 2022 cell segmentation challenge (NeurIPS CellSeg)\footnote{The online challenge can be accessed via \url{https://neurips22-cellseg.grand-challenge.org}.}.
Experiment results show that the cell segmentation model learned via MMCS is capable of coping with the diverse resolution of microscopy images and multiple cell shapes.

The rest of this paper is organized as follows:
Section 2 reviews related works.
Section 3 introduces the proposed MMCS method, which includes implementation details for the nucleus-assisted global recognition, self-adaptive diameter filte and the temporal-ensembling models procedures.
Section 4 presents implementation details of our experiments.
Section 5 concludes the paper.

\section{Related Works}
Cell segmentation algorithms can generally be categorized based on three criteria: the principle of cell detection, the computed image features, and the segmentation method itself, while state-of-the-art methods often combine various strategies to achieve better results \cite{ulman2017objective}. The principle includes finding uniform areas, boundaries \cite{jiang2006novel,kevin2008general}, or bright spots and maxima \cite{Wuttisarnwattana2016AutomaticSC}. Image features can range from simple pixel intensities to complex descriptors of shapes or textures. Segmentation methods range from simple thresholding \cite{Lerner2001AutomaticSC,Chen2006AutomatedSC}, hysteresis thresholding \cite{Henry2013PhagoSightAO}, shape matching \cite{Cicconet2013WaveletbasedCH,Tretken2017NetworkFI}, and edge detection \cite{Whlby2004CombiningIE}, to sophisticated approaches like machine learning \cite{ronneberger2015u,Schiegg2015GraphicalMF} and energy minimization \cite{Bensch2015CellSA,mavska2013segmentation,Dzyubachyk2010AdvancedLC}.

\subsection{Cell Segmentation in Cell Tracking Challenge}
Cell-tracking algorithms are designed to identify individual cells and follow them over time to gain biological insights from time-lapse microscopy recordings of cell behavior \cite{ulman2017objective}. The Cell Tracking Challenge (CTC) \footnote{http://celltrackingchallenge.net/} held by the IEEE International Symposium on Biomedical Imaging (ISBI) is aimed to promote the development and objective evaluation of cell segmentation and tracking algorithms. Since the sixth edition, the primary focus of CTC has been put on methods exhibiting better generalizability. The CTC is a powerful resource for cell segmentation algorithms. The segmentation methodologies used in the competition include thresholding, energy minimization, machine learning, region growing, and so on \cite{ulman2017objective}. The comparison of the competing algorithms shows that most practical scenarios tracking by detection methods outperformed tracking by contour evolution methods. And also, algorithms using modern machine-learning approaches perform best in most segmentation scenarios. 
\subsection{Deep Learning-based Cell Segmentation}
Recently, more and more studies have leveraged deep learning techniques to realize cell segmentation.
At first, the approaches of deep learning networks are based on Mask R-CNN architecture \cite{tsai2019usiigaci}, which are stain-free, instance-aware segmentations. Usiigaci, an all-in-one, semi-automated pipeline can segment, track, and visualize cell movement and morphological changes in phase contrast microscopy \cite{tsai2019usiigaci}. Usiigaci has a good performance on electrotaxis of NIH/3T3 fibroblasts. 
Recently more algorithms such as StarDist \cite{schmidt2018cell}, MiSiC \cite{panigrahi2021misic}, Omnipose \cite{cutler2022omnipose} and Cellpose \cite{stringer2021cellpose} are based on the U-Net architecture. The network U-Net can be trained end-to-end from very few images and outperforms most of the prior methods owed to a large number of feature frames in the upsampling part \cite{ronneberger2015u,zhou2018unet++,falk2019u}. StarDist, localizing cell nuclei via Star-convex polygons, performs well on a challenging dataset of fluorescence microscopy images with very crowded nuclei \cite{Schmidt2018CellDW}.
MiSiC can segment single bacteria from interacting bacterial communities in complex images at very high throughput \cite{panigrahi2021misic}. Omnipose, a segmentation algorithm based on Cellpose which is introduced in the Methodology part, excellently performs on mixed bacterial cultures, antibiotic-treated cells and cells of elongated or branched morphology \cite{cutler2022omnipose}.

\section{Methodology}
\label{sec:formatting}

Suppose $L=\{(x^U_i,y_i)\}$ is a labeled training dataset, where each sample point $x^U_i \in \mathcal X$ is assigned with a ground truth label $y_i \in \mathcal Y$.
Also, suppose $U = \{x^U_i\}$ is an unlabeled training dataset, where $x^U_i$ is the $i$-th unlabeled sample point.
Then, the goal of semi-supervised cell segmentation is to learn a cell segmentation model $f_\theta : \mathcal X \rightarrow \mathcal Y$ from both the labeled dataset $L$ and unlabeled dataset $U$, where $\theta \in \Theta$ is the learned model parameter.

In this section, we first revisit the existing Cellpose method, illustrate its disadvantages in segmenting some specific types of cell images, and then introduce our proposed MMCS cell segmentation approach in detail.

\subsection{Cellpose Revisiting}
Cellpose \cite{stringer2021cellpose} is a deep learning-based segmentation method, which amassed tons of varied images of nuclei from many different laboratories around the world. 
It is proposed to segment types of cells without requiring parameter adjustments, new training data or further model retraining.
Different from classical segmentation approaches that usually adopt the grayscale values of images to create topological maps, Cellpose leverages an intermediate image representation that is constructed via a process of simulated diffusion to form smoother topological maps.
As a result, compared with previous methods, Cellpose can perform better in segmenting many types of cells that are not blob-like.

\begin{figure*}
  \centering
  \begin{subfigure}{\linewidth}
      \centering
      \includegraphics[height=0.19\linewidth, width=0.2\linewidth]{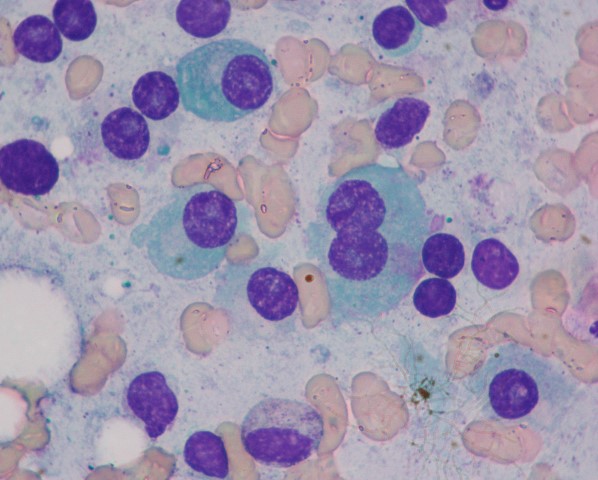}
      \hspace{0.5em}
      \includegraphics[height=0.19\linewidth, width=0.2\linewidth]{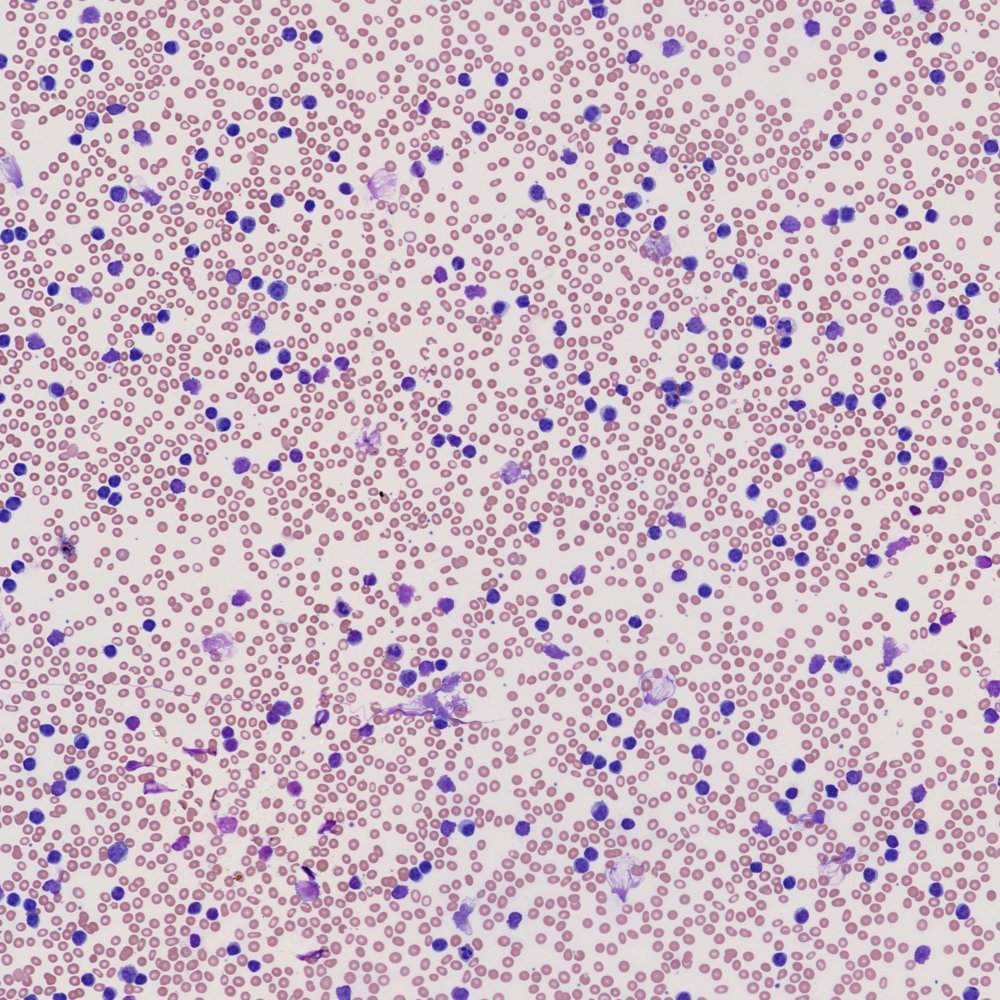}
      \hspace{0.5em}
      \includegraphics[height=0.19\linewidth, width=0.2\linewidth]{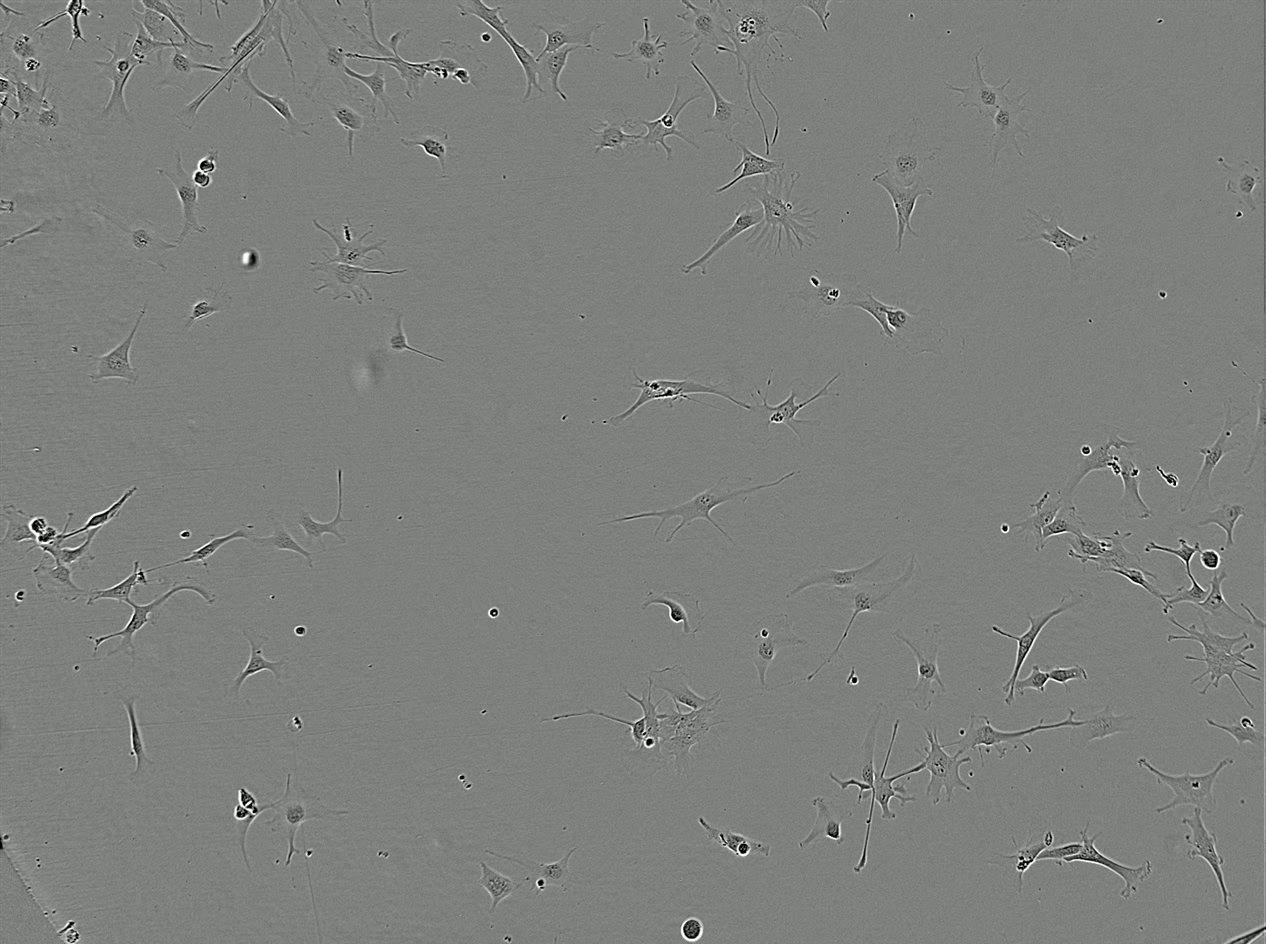}
      \hspace{0.5em}
      \includegraphics[height=0.19\linewidth, width=0.2\linewidth]{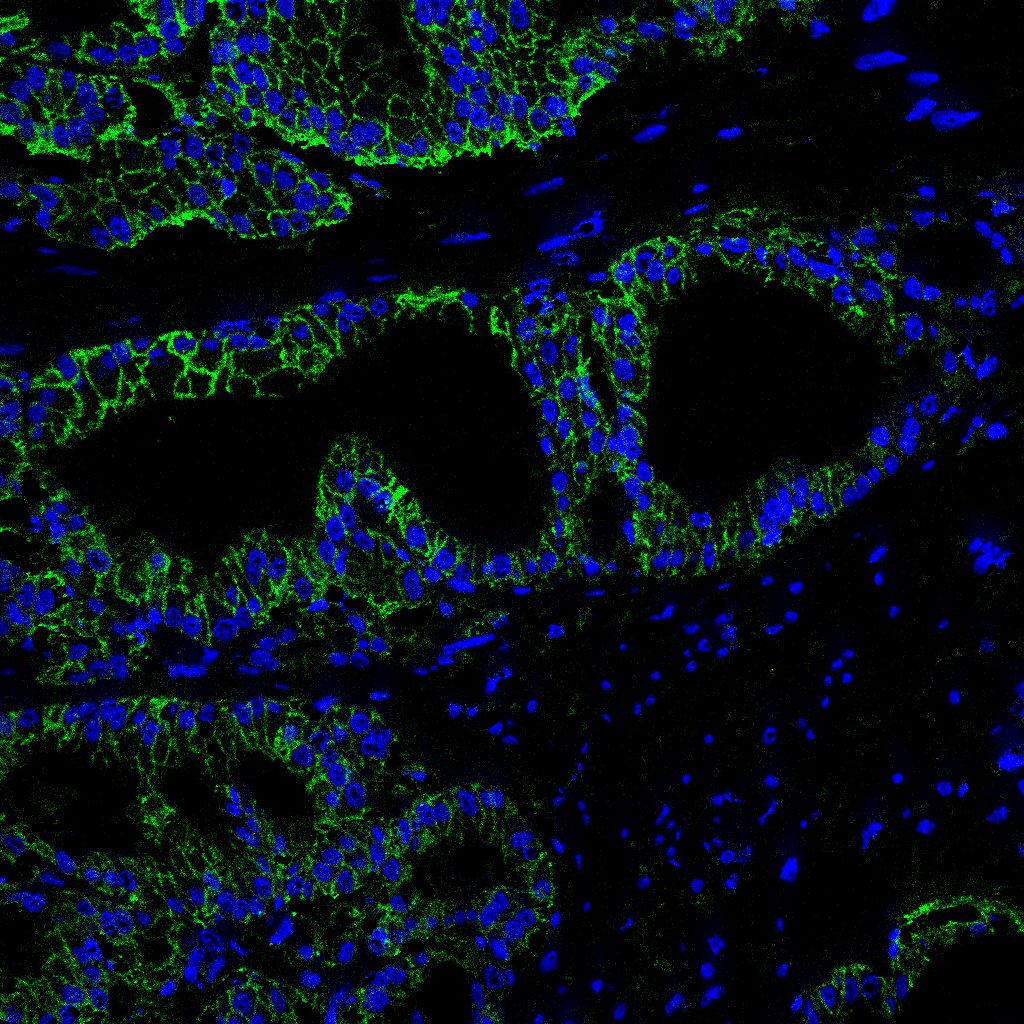}
      \subcaption{Original cell images.}
  \end{subfigure}
  
  \begin{subfigure}{\linewidth}
      \centering
      \includegraphics[height=0.19\linewidth, width=0.2\linewidth]{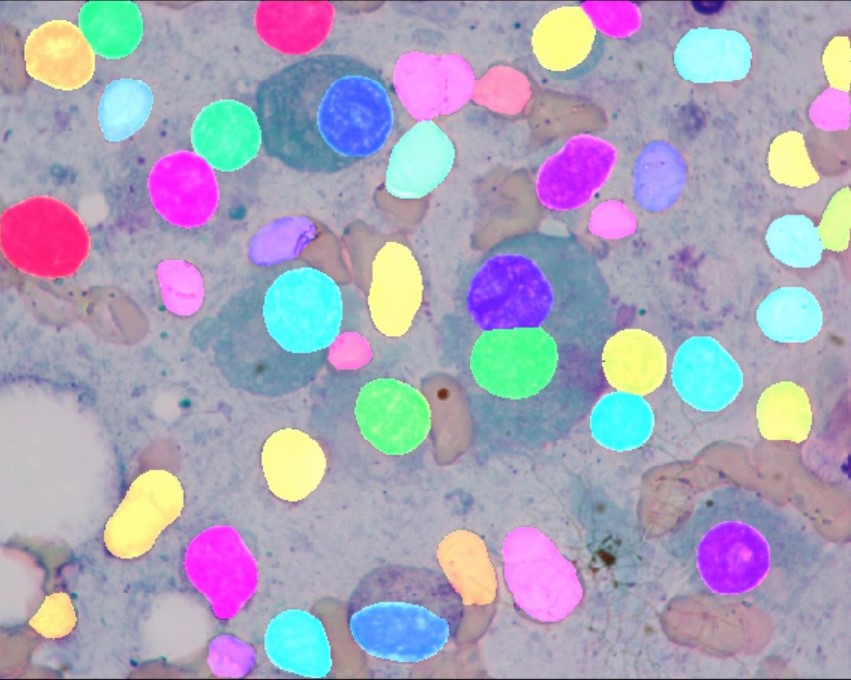}
      \hspace{0.5em}
      \includegraphics[height=0.19\linewidth, width=0.2\linewidth]{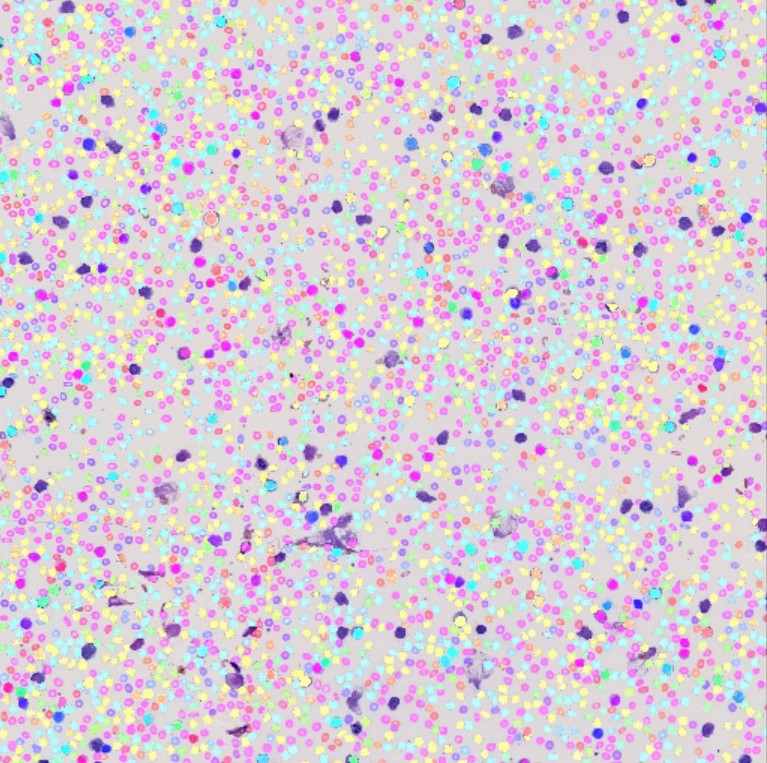}
      \hspace{0.5em}
      \includegraphics[height=0.19\linewidth, width=0.2\linewidth]{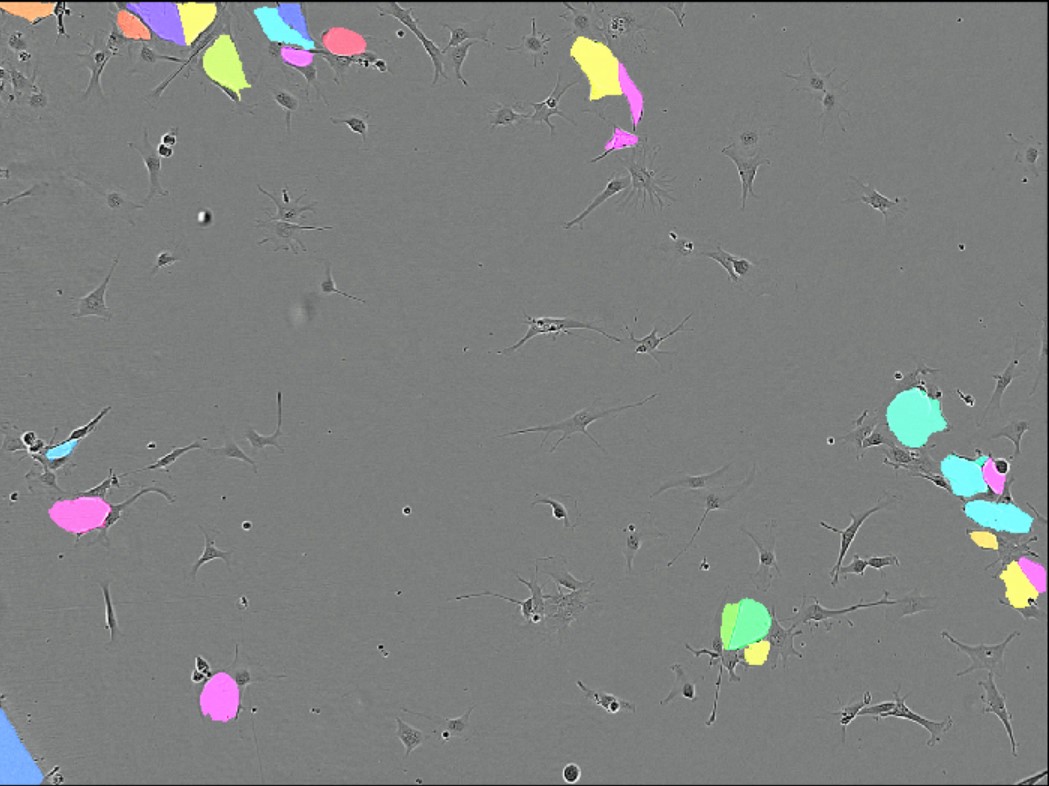}
      \hspace{0.5em}
      \includegraphics[height=0.19\linewidth, width=0.2\linewidth]{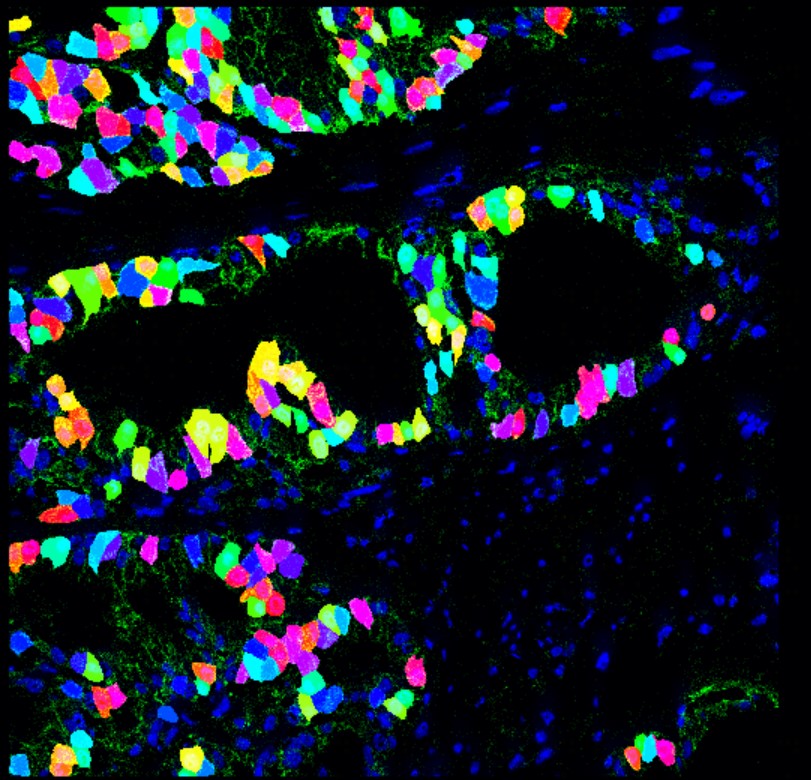}
      \subcaption{Cell segmentation processed by Cellpose.}
  \end{subfigure}
  
  \begin{subfigure}{\linewidth}
      \centering
      \includegraphics[height=0.19\linewidth, width=0.20\linewidth]{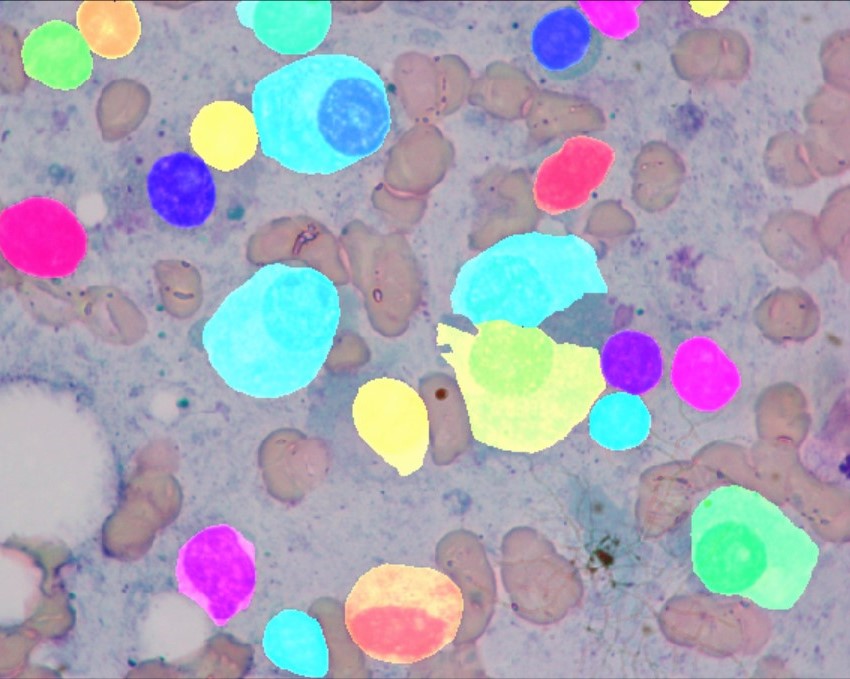}
      \hspace{0.5em}
      \includegraphics[height=0.19\linewidth, width=0.20\linewidth]{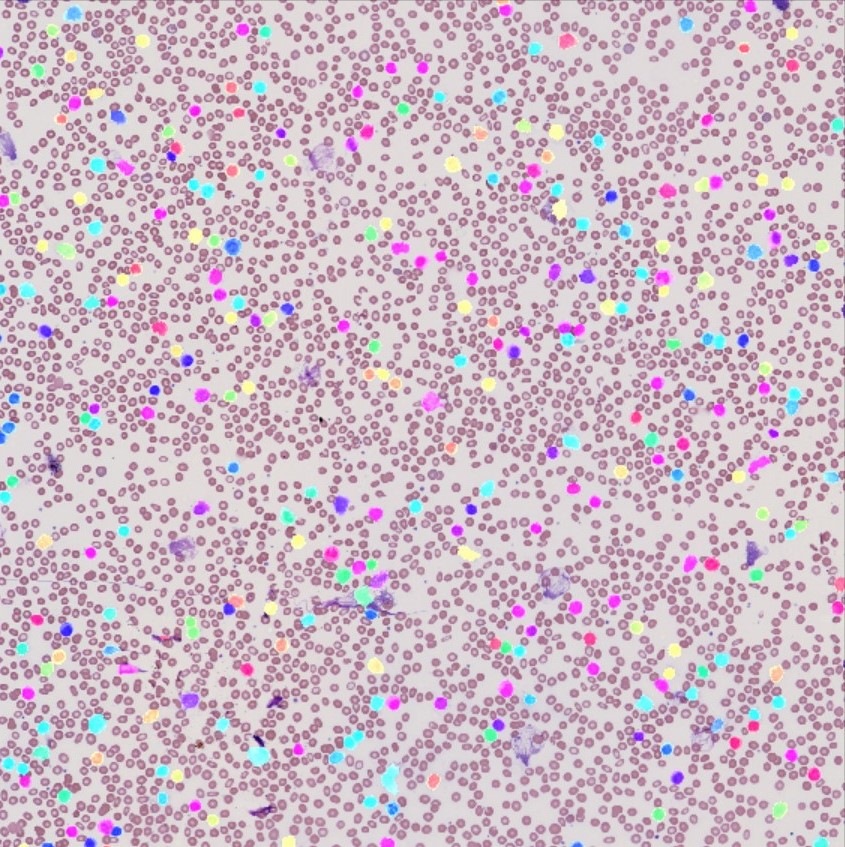}
      \hspace{0.5em}
      \includegraphics[height=0.19\linewidth, width=0.20\linewidth]{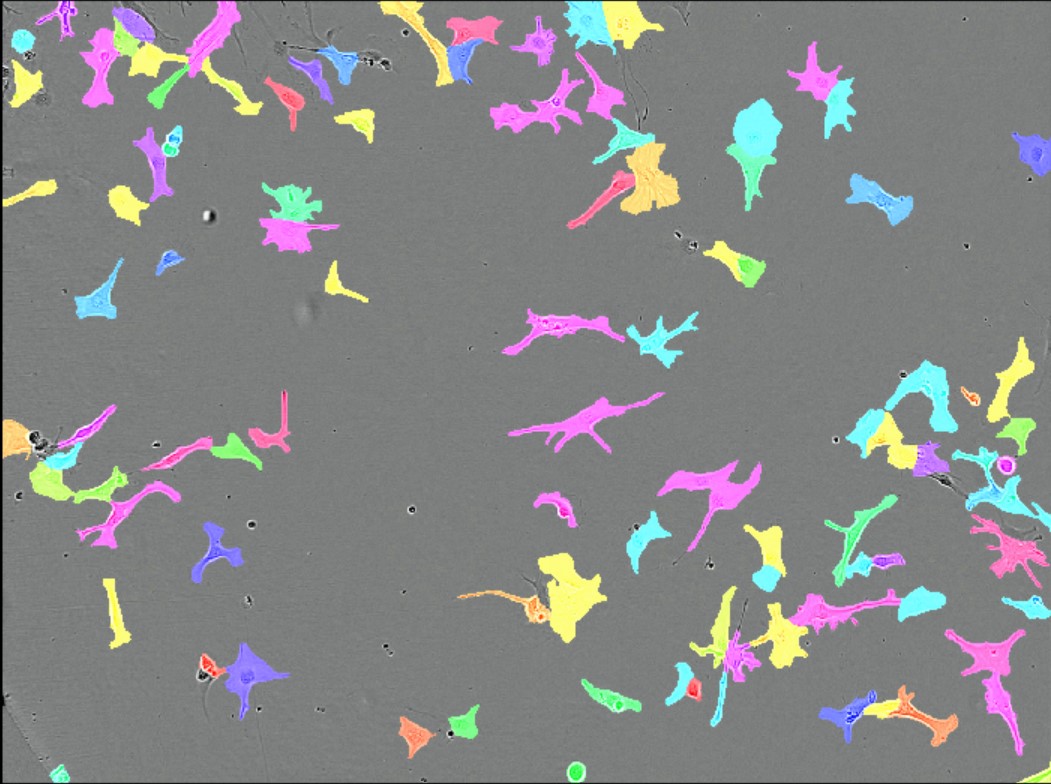}
      \hspace{0.5em}
      \includegraphics[height=0.19\linewidth, width=0.20\linewidth]{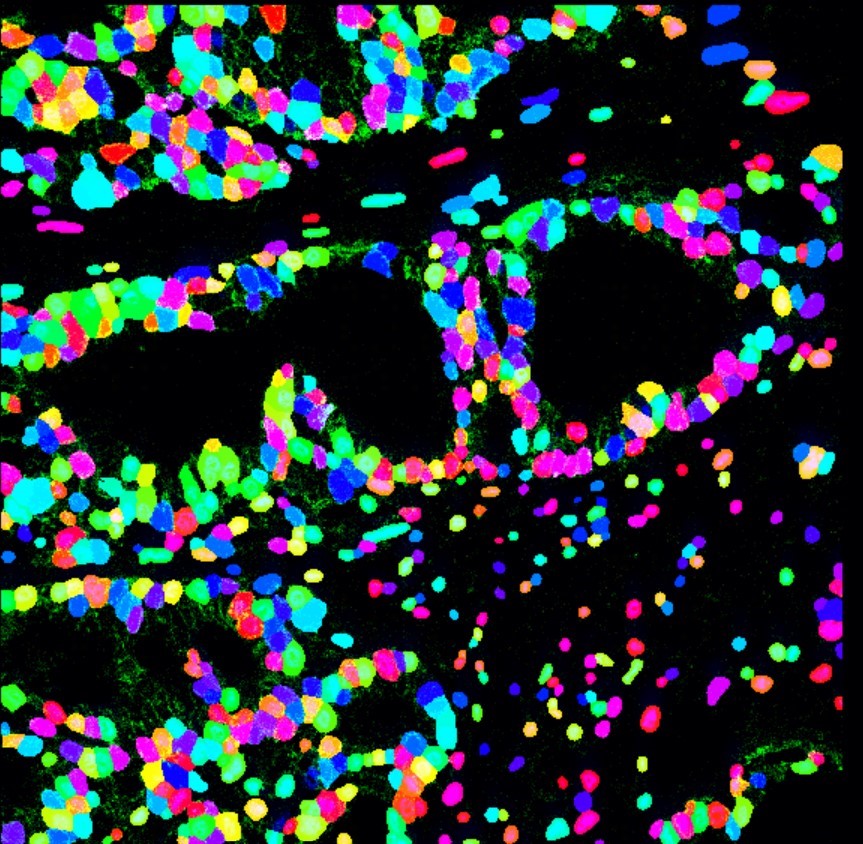}
      \subcaption{Cell segmentation processed by MMCS.}
  \end{subfigure}
  \caption{Example results of Cellpose and MMCS \cite{stringer2021cellpose}.
  (a) Representative original images of stained cells (purple) mixed with weak stained tissues (pale pink) (left two columns), light microscopy images with cells of elongated or branched morphology (the third column), and fluorescent images (the fourth column). (b) Predicted masks of original images in (a), using Cellpose with cyto2 pre-trained model. (c) Masks of original images in (a) predicted utilizing MMCS. Cellpose cannot distinguish between stained cells and weak stained tissues, and it is challenging to segment cells with elongated or branched morphology in light microscopy images. In fluorescent images, Cellpose segmentation missed a lot of cells.}
  \label{fig:cellpose}
\end{figure*}

However, we find that Cellpose could not perform cell segmentation well on some specific types of cell images such as immunohistochemistry images and light microscopy images.
For example, in our experiment, stained cells mixed with weak stained tissues and cells of elongated or branched morphology can not be segmented well using Cellpose, as shown in Fig.~\ref{fig:cellpose} (b).
This is likely because of that Cellpose simplified the cell segmentation process into detecting cell membranes boundary and as a result, failed to leverage domain-specific knowledge such as cellular structures or multi-microscopic cell views to promote the segmentation.
Besides, the performance of Cellpose highly relied on the quality of training data and the quantity of annotated data, but it is usually difficult to obtain high-quality, annotated data in real-world applications.


\subsection{Overview Architecture of MMCS}

To tackle the challenge of limited high-quality annotated cell images and enhance the performance of cell segmentation, this paper proposed Multi-Microscopic-view Cell semi-supervised Segmentation (MMCS) to effectively learn cell segmentation model from less images of mixed cells from different microscopic views.
MMCS consists of three key components:
(1) nucleus-assisted global recognition,
(2) self-adaptive multi-scale cell filtering,
and (3) temporal model ensembling with loss reweighting training.
The overall workflow of MMCS is shown in Fig.~\ref{fig:mmcs-workflow}.

Specifically, MMCS employs an extra nucleus channel that incorporates significant cellular features to enhance the learning model.
In general, it can simultaneously incorporate nucleus knowledge into global recognition patterns, facilitating comprehensive decision-making under fuzzy recognition scenarios, such as when cells are overlapping.
Then, MMCS assigns a personalized self-adaptive cell diameter filter to each cell subset, enabling the extraction of knowledge from training data of all cell sizes, and adapting to various cell sizes during inference.
Finally, MMCS utilizes a temporal-ensembling model to improve the stability of semi-supervised training.
To ensure the accuracy and reliability of the pseudo labels generated during semi-supervised learning, we adopt a pre-trained model that is trained on all available labeled data to initialize pseudo labels for unlabeled data.
The pseudo labels will then be updated with the temporal-ensemble training model along the semi-supervised learning process.
This approach enables the model to leverage both labeled and unlabeled data in a more effective way, ultimately improving the accuracy and robustness of the trained model.
Furthermore, the weight of unlabeled loss contributing to the total loss is selected to optimize for MMCS. Specifically, we design the following objective loss function for MMCS.
\begin{gather}
    \mathcal{L}
    =  (1-w)\mathcal{L}(\theta, L , \rho) + w \mathcal{L}(\theta, U , \rho),
\end{gather}
where $\mathcal{L}$ is a cross-entropy loss function as that used in \cite{stringer2021cellpose}, $w > 0$ is a preset importance coefficient for reweighting unlabeled dataset $U$, $L$ is the labeled dataset and $\rho$ is the adaptive diameter of the cell filter.


\subsection{Nucleus-assisted Global Recognition}
Two-channel cell images contain the primary channel corresponding to the cytoplasmic label, and the optional second channel corresponding to the nucleus, which in all cases is a  4,6-diamidino-2-phenylindole- (DAPI-) stain appearing blue in color. The second nucleus channel can provide cell localization information. MMCS employs an extra nucleus channel to leverage dominant cytoarchitecture to express relative location information and cell characteristic. Technically, after augmentation, the green cytoplasmic and blue nuclei channels are sliced and input into the neural network. When the second nucleus channel is not available, it is replaced with an image of zeros. In this way, it assists to improve the global recognition decision-making, rather than framing the cell-seg task as a general compute vision segmentation task, where models learning from the naive edge detection lack domain knowledge, and it can not effectively identify a single cell when cells aggregate heavily. 
\begin{figure*}[h]
  \centering
  \includegraphics[width=\linewidth]{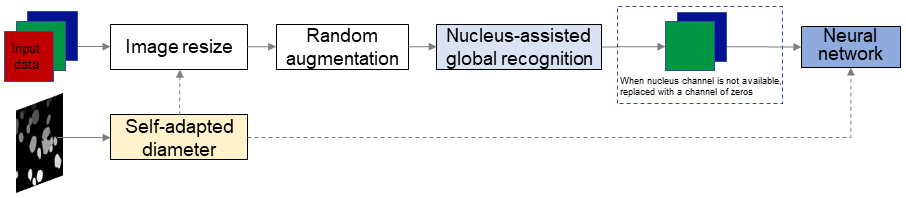}
  \caption{Overview of nucleus-assisted global recognition and self-adaptive diameter filter. The cell self-adaptive diameters are calculated by analyzing label areas. Before performing augmentations, we resize the training data according to the self-adaptive diameters. And then, nucleus-assisted global recognition is performed, leading to two channels of images, green and blue, input into the neural network. When the second blue channel is unavailable, it is replaced with an image of zeros. Additionally, the mean cell self-adaptive diameter is saved into the network.}
  \label{fig:NGR-workflow}
\end{figure*}

\subsection{Self-adaptive Diameter Filter}
Cells in biology exhibit diversity not only in their types but also in their shapes and characteristics. This includes their size, shape, color, internal structures, and functions. Additionally, different types of cells may also exhibit significant differences in their morphologies and features. For instance, red blood cells are round with a flattened central area that can accommodate oxygen and carbon dioxide, while neurons typically have a star-shaped morphology with multiple branching processes for transmitting electrical signals. To incorporate the maximum possible information about cell morphology,  MMCS assigns a personalized self-adaptive cell diameter filter to each cell subset. The cell diameters are calculated by analyzing label areas, which guide the training data to resize before performing augmentations and then saved into the training model as the mean diameter of training regions of interest (ROIs).  During inference, an argument vector called \textit{\textbf{diameter}} is selected and set to the mean cell diameter of the inference images. This guides the resizing of the inference ROIs to match the mean diameter of the training ROIs of the model.
A simplified framework of nucleus-assisted global recognition and self-adaptive diameter filter is summarized as Fig.\ref{fig:NGR-workflow}.


\subsection{Temporal-Ensembling Models for Semi-supervised Training}
Since there is a scarcity of annotated medical images, models based on the concept of semi-supervised learning are extensively employed for medical image segmentation \cite{Zhang2017DeepAN,Uncertainty2021Cao}. Semi-supervised learning allows models to leverage unlabeled data without any restrictions, greatly reducing their reliance on labeled data and accurately identifying the model's generalization direction. As a result, these models have a higher potential for practical applications. A key factor determining the performance of semi-supervised segmentation is to ensure the accuracy and reliability of pseudo labels of unlabeled data.

In MMCS, we leverage a temporal-ensembling approach to improve the performance of semi-supervised cell segmentation.
The overall procedures are presented as Algorithm~\ref{algo:temporal_ensembling}.
Specifically, we first train the model on the small labeled dataset and use the trained model to predict and assign an initial pseudo label $\widetilde{z}_0$ to the unlabeled data.
We then perform semi-supervised learning to fine-tune the model on both the labeled and unlabeled datasets with the help of pseudo labels.
It is worth noting that the pseudo labels will be updated every $T$ epochs as below,
\begin{gather}
    \widetilde{z}_k \leftarrow \frac{1}{2} (\widetilde{z}_{k-1} + f_{\theta_{kT}}(x)),
\end{gather}
where $\widetilde{z}_k$ $f_{\theta_{kT}}$ is the pseudo label and the model obtained at the ($kT$)-th training iteration.
Throughout this paper, the epochs number $T$ is set as $100$.

\begin{algorithm*}
\caption{Pseudo code of the Temporal-ensembling model}
\label{algo:temporal_ensembling}
\begin{algorithmic}[1]
\Require: {
Labeled training dataset $L$, \\
unlabeled training dataset $U$, \\
semi-supervised training epochs number $E$, \\
pseudo-label update interval $T$, \\
model $f_{\theta_0}$ that is pretrained on the labeled dataset $L$, \\
and loss-reweighting factor $w$}.
 \State Initialize the pseudo-label of $U$ as $\tilde Z^U \leftarrow \mathrm{Concatenate}_{x \in U}[f_{\theta_0}(x)]$
 \For {$k$ in $1$ to $E$}
 \State{$\theta_k \leftarrow \theta_{k-1}$}
 \For {each minibatch $B$ from the overall dataset}
 \State {Seperate the minibatch into labeled data $B^L = \{(x^L_i, y_i)\}$ 
 \Statex \quad \quad  \quad \quad  \quad \quad \quad \quad \quad \quad \quad \quad \quad \quad  and pseudo-labeled data $B^U = \{(x^U_i, \tilde z_i)\}$}
 \State{ $\theta_k \leftarrow \theta_k - \nabla_\theta \left\{ \frac{1-w}{|B^L|}\sum_{(x^L_i, y_i) \in B^L} \mathcal{L}(f_{\theta_k}(x^L_i), y_i) + \frac{w}{|B^U|}\sum_{(x^U_i, \tilde z_i) \in B^U} \mathcal{L}(f_{\theta_k}(x^U_i, \tilde z_i)) \right\}$ }
 \EndFor
 \If{$k \bmod T$ is $0$}
  \State Update the pseudo-label of $U$ as $\tilde Z^U \leftarrow \frac{1}{2} (\tilde Z^U + \mathrm{Concatenate}_{x \in U}[f_{\theta_k}(x)])$
 
  \EndIf
 \EndFor
 \Return{$\theta_E$}
\end{algorithmic}
\end{algorithm*}

\section{Experiments}
\label{sec:formatting}
\subsection{Implementation Details}
We trained the proposed MMCS model on a multi-microscopic views dataset from NeurIPS CellSeg combined with the Cellpose and Omnipose datasets. The image appearances are diverse because of different tissues and staining methods. Metadata (e.g., modality, tissue) for each image will not be provided. Therefore, the model is necessitated to be robust to different modalities, tissues, and staining methods. 
MMCS utilized CPnet, the model architecture in Cellpose, as its own model structure, and adopted the gradient flow tracking in CellPose for cell masks \cite{stringer2021cellpose}. For data augmentation, the self-adaptive diameter filter determined how much to resize images before performing other augmentations, including random rotation, normalization and affine transformation. 
We used the SGD optimizer with an initial learning rate of 0.1, momentum of 0.9, and weight decay of 0.00001. The learning rate was linearly increased from zero to 0.1 over the first ten epochs. A learning rate annealing schedule was also selected to minimize training set loss. The batch size for all models was 16 or 32. During the semi-training phase, the training loss was divided into \{$supervised\_loss,  unsupervised\_loss$\} and their corresponding factors were multiplied to guide learning contribution. The model was trained for 2000-4000 iterations with no specific stopping criterion and validated every 1000 iterations.  
We validated the MMCS on 101 images including one whole-slide image (10,000x10,000) on NeurIPS CellSeg. See Supplemental Material for further details on the development environment and training hyperparameters.


\begin{table}[t]
  \centering
  \begin{tabular}{@{}lccc@{}}
  \toprule
  \multirow{2}{2cm}{Datasets} & \multicolumn{2}{c}{Prediction Model} \\
  \cmidrule{2-3}
    & phase1& phase2\\ 
  \midrule
    dataset1  & 0.7729 & 0.7944 \\
   +dataset2 &  0.8117 & 0.8239 \\
    \bottomrule
  \end{tabular}
  \caption{MMCS validation F1 scores by different models trained with different datasets in pre-trained model training and semi-supervised training phase.  The left row shows the F1 scores in the pre-training phase (phase1). The right row is the results in the semi-supervised training phase (phase2).  "dataset1" denotes the labeled data from NeurIPS CellSeg, and "+dataset2" denotes combined with Cellpose and Omnipose datasets. In semi-supervised training phase2, the unlabeled data from NeurIPS CellSeg were also used.}
  \label{tab:result-1}
\end{table}

\subsection{Datasets}
Our experiments adopt one dataset from the NeurIPS CellSeg and two public datasets as follows:

 \textbf{NeurIPS CellSeg}: contains 1000 labeled and more than 1500 unlabeled image patches from various microscopy types, tissue types, and staining types. The labeled training set consists of four microscopy modalities, including Brightfield (300 patches), Fluorescent (300 patches), Phase-contrast (200 patches), and Differential interference contrast (200 patches).

\textbf{Cellpose} \cite{stringer2021cellpose}: includes fluorescent cell images, images of cells from brightfield microscopy and a small set of nonmicroscopy images. We used 539 images of the train set.

\textbf{Omnipose} \cite{cutler2022omnipose}: we used 501 images of the train set containing 14 bacterial species, such as Escherichia coli, Shigella flexneri, Acinetobacter baylyi, E. coli and so on. 
\begin{table}[t]
  \centering
  \begin{tabular}{@{}cccc@{}}
    \toprule
  \multicolumn{4}{c}{F1 Score on the Cellpose Dataset} \\
  MMCS (Ours) & Cellpose & Stardist & Mask R-CNN \\ 
    \midrule
  \textbf{0.7732}  & 0.77&0.61&0.61 \\
    \bottomrule
  \end{tabular}
  \caption{
    The F1-Score results of different cell segmentation approach on the Cellpose dataset.
    A higher F1-Score indicates a better cell segmentation performance.
    The results suggest that our MMCS outperforms existing approaches.
  }
  \label{tab:result-8}
\end{table}

\begin{table}[t]
  \centering
  \begin{tabular}{@{}cc|c|c|c@{}}
    \toprule
  \multicolumn{2}{c|}{Pretrained Model} & \multirow{2}{2.1cm}{Add nucleus channel} & \multirow{2}{1.5cm}{Inference diameter} &\multirow{2}{1.5cm}{F1 score}\\
    cyto2 & cyto & & \\ 
    \midrule
    \checkmark &×&×&30 pixels&0.5815\\
    \checkmark &×&\checkmark&30 pixels&\textbf{0.7397} \\
    ×&\checkmark&\checkmark&30 pixels&0.7342 \\
    ×&×&\checkmark&30 pixels&0.7342 \\
    \bottomrule
  \end{tabular}
  \caption{Comparison of training with or without aiding nucleus channel. MMCS outperformed with aiding blue nucleus channel and cyto2 pretrained model. \checkmark and ×  were used to denote whether or not the corresponding module is enabled during the experiments. × in "Pretrained model" row means no use for pretrained models, and × in "Add nucleus channel" means  no adding nucleus channel.}
  \label{tab:result-2}
\end{table}

\subsection{Cell Segmentation Performance of MMCS}
The performance is evaluated via the F1-score at the IoU threshold of 0.5 for true positives. The results of our models from each phase can be found in Table~\ref{tab:result-1} and the visualization of MMCS prediction results can be found in Fig.~\ref{fig:cellpose} (c). With temporal-ensembling in semi-supervised training using a learning rate decay and weighted loss, our MMCS model achieves F1-score 0.8239 on validation datasets with 101 images. The results showed that MMCS is capable of coping with the diverse resolution of microscopy images and multiple cell shapes.


\subsection{Comparison with Previous Methods}
We conduct the comparison experiments with state-of-the-art algorithms in  Table~\ref{tab:result-8}.

\textbf{Results on Cellpose generalized test dataset} As shown in  Table~\ref{tab:result-8}, on the Cellpose generalized test set, the Cellpose, Stardist and Mask R-CNN model have average precisions of 0.77, 0.61 and 0.61 respectively \cite{stringer2021cellpose}. MMCS achieve an average precision of 0.7732, significantly outperforming Stardist and Mask R-CNN and comparable to Cellpose, even slightly better.

\textbf{Leaderboad of NeurIPS CellSeg} The outstanding winner of NeurIPS CellSeg achieves a F1 score of 0.9070 using a MEDIAR framework that harmonizes data-centric and model-centric approaches as the learning and inference strategies \cite{Lee2022MEDIARHO}. To train this model, the team utilized four external public datasets containing 7,242 labeled data. They also examined semi-supervised learning approaches to exploit the unlabeled images, unfortunately with no improvement of performance. MMCS ranked the 17th place on the tuning set in NeurIPS CellSeg with fewer labeled data, expecting that semi-supervised learning approaches can be a promising alternative direction for using unlabeled images.


\begin{table}[t]
  \centering
  \begin{tabular}{@{}ccc@{}}
    \toprule
 Dataset &  Diameter  & F1 score  \\
    \midrule
  \multirow{7}{1cm}{dataset1}&30 pixels & 0.7397 \\
  &35 pixels & 0.7650 \\
  &40 pixels & \textbf{0.7729} \\
  &45 pixels & 0.7485 \\
  &50 pixels & 0.7412 \\
  &55 pixels & 0.7078 \\
  &60 pixels & 0.6661 \\
  \midrule
  +dataset2 &40 pixels&0.8117 \\
    \bottomrule
  \end{tabular}
  \caption{Comparison of different diameters on MMCS during inference. MMCS achieved a better F1-score when the diameter was set to 40 pixels. "dataset1" and "+dataset2" denote labeled dataset of NerIPS CellSeg and combined with Cellpose and Omnipose datasets.}
  \label{tab:result-3}
\end{table}

\begin{table}[t]
  \centering
  \begin{tabular}{@{}ccc@{}}
\toprule
  \multirow{2}{1cm}{\textit{w}} & \multicolumn{2}{c}{F1 score} \\
  \cmidrule{2-3}
    & dataset1& +dataset2\\ 
   \midrule
    0.05 &0.7684 & \textbf{0.8188} \\ 
    0.1& 0.7472 &  0.7915\\
    0.2 & 0.7557 &0.7971 \\ 
    0.3 &0.7668  & 0.7680 \\
    0.4 &\textbf{0.7904} & 0.7621  \\
    0.5 & 0.7850 &0.7409   \\
    0.6 & 0.7371 & 0.7452 \\
    0.7  & 0.6878  & 0.7102\\
    \bottomrule
  \end{tabular}
  \caption{Comparison of the effects of different alpha factor on MMCS. On labeled and unlabeled datasets of NerIPS CellSeg, MMCS achieved a better F1-score when \textit{\textbf{w}} is 0.4 compared with other weight importance. The learning rate is set to a fix 0.1 value (middle row). When adding Cellpose and Omnipose datasets, with a weight importance of 0.05, MMCS performed better. Learning rate dacay strategy is used (right row). }
  \label{tab:result-4}
\end{table}

\subsection{Ablation Study}
\textbf {Nucleus-assisted Global Recognition} To examine the effect of nucleus-assisted global recognition, green and blue channels of the two-channel images were input into the neural network. When the second blue channel was unavailable, it was replaced with an image of zeros. The result in Table~\ref{tab:result-2} showed that compared with one green channel input, with aiding blue nucleus channel and cyto2 pretrained model, MMCS achieved the best 0.7397 F1-score on the labeled images from NeurIPS CellSeg. The inference diameter was set to default 30 pixels.

\textbf{Self-adaptive Diameter Filter}   To further investigated the effect of self-adaptive diameter filter during MMCS inference, we used an argument vector \textit{\textbf{diameter}} guiding the inference ROIs to resize to the size of the mean diameter of training ROIs.
As shown in Table~\ref{tab:result-3}, MMCS achieved a better 0.7729 F1-score on the labeled images from the NeurIPS CellSeg dataset when the diameter was set to 40 pixels.
When adding the Cellpose and Omnipose datasets, the F1-score achieved 0.8117. 

\textbf{Temporal-ensembling Models} 
In the semi-supervised training phase, we exploit the 1,500+ unlabeled images in the multi-microscopic views train datasets from NeurIPS CellSeg using the temporal-ensembling approach. The pseudo labels of unlabeled images were obtained from pre-trained model predictions accumulating previous network predictions every 100 epochs. The loss is weighted by two components, the cross-entropy loss evaluated for labeled inputs and unlabeled inputs. The unlabeled input loss was calculated between the pseudo labels and the current network predictions. Results in  Table~\ref{tab:result-4} show the effect of different weighted loss. The coefficient \textit{\textbf{w}} is the weight ratio of unlabeled loss. The results show that MMCS achieved a better F1-score of 0.7904 when \textit{\textbf{w}} was 0.4 on the multi-microscopic views dataset. When adding the Cellpose and Omnipose datasets,  MMCS outperformed with a 0.05 weight coefficient. In addition, we observed that MMCS converged at epoch 2000 and performed poorer at epoch 3000 and 4000 when the learning rate is set to 0.1 as shown in  Table~\ref{tab:result-5}. To minimize training set loss, the learning rate annealing schedule was optimized with the results shown in the next paragraph.

 \begin{table}[t]
  \centering
  \begin{tabular}{@{}cccc@{}}
    \toprule
  \textit{w}&Epoch&Decay schedule&F1 score \\
    \midrule
    \multirow{4}{1cm}{0.4}& 1000  &×& 0.7510  \\
     & 2000 &×& \textbf{0.7904}  \\
     & 3000&×& 0.7721   \\
     & 4000&×&0.7513  \\ 
     \midrule
    0.4&3000&\checkmark&\textbf{0.7944} \\
    0.2&3000&\checkmark&0.7882  \\
    0.05&3000&\checkmark&0.7686  \\
    \bottomrule
  \end{tabular}
  \caption{Comparison of the effects on MMCS with or without learning rate annealing schedule on the datasets from NeurIPS CellSeg. MMCS achieved a better performance with a learning rate annealing schedule. \checkmark in "Decay schedule" row means using learning rate annealing schedule.}
  \label{tab:result-5}
\end{table}

\begin{table}[t]
  \centering
  \begin{tabular}{@{}cccc@{}}
    \toprule
  \textit{w}&Epoch&Starting epoch&F1 score \\
    \midrule
    \multirow{4}{1cm}{0.05}& 1000  &2000& 0.7971  \\
     & 2000 &2000& \textbf{0.8188}   \\
     & 3000 &2000&0.8092    \\
     & 4000&2000&0.8170  \\ 
    \midrule
    0.05&2000&1000&0.8162 \\
    \midrule
     0.05&2500&1500&\textbf{0.8239}  \\
    
    \bottomrule
  \end{tabular}
  \caption{Comparison of the effects of different learning rate annealing schedule on MMCS on the datasets from NeurIPS CellSeg combined with the Cellpose and Omnipose datasets. MMCS achieved a best performance with learning rate annealing schedule starting at at epoch 1500."starting epoch" denotes the epoch at which  the learning rate annealing schedule starts.}
  \label{tab:result-6}
\end{table}

\textbf{Optimization of the Framework}  
To optimize the framework, the learning rate was chosen to minimize training set loss. The learning rate annealing schedule was first started at epoch 2000, halving the learning rate every hundred epochs to a minimum of  0.0016. With this learning rate annealing schedule, MMCS achieved a better performance of 0.7944 at epoch 3000 on the multi-microscopic views datasets from NeurIPS CellSeg, shown in  Table~\ref{tab:result-5}. However, when adding the Cellpose and Omnipose labeled train datasets, MMCS achieved a poor performance of 0.8092 at epoch 3000 compared to the 0.8188 F1 score at epoch 2000, shown in  Table~\ref{tab:result-6}. That indicates that with this learning rate annealing schedule, the framework did not achieve the minimization of the training set loss with more labeled data. Therefore, the learning rate halving schedule was set to start at epoch 1000 or 1500 with a total training epoch of 2000 or 2500. The results in Table~\ref{tab:result-6} showed that  MMCS achieved a better 0.8239 F1 score when the learning rate halving schedule started at epoch 1500. 

\begin{table}[t]
  \centering
  \begin{tabular}{@{}lc@{}}
    \toprule
  Architecture  & F1 score  \\
    \midrule
  U-Net & 0.5353  \\
  Swinunetr & 0.5603 \\ 
  MMCS & \textbf{0.8239} \\
    \bottomrule
  \end{tabular}
  \caption{Comparison between MMCS with two CNN models, U-Net and Swinunetr. MMCS significantly outperformed U-Net and Swinunetr}
  \label{tab:result-7}
\end{table}

\textbf{Model Structure}
As shown in Table~\ref{tab:result-7}, we compared our MMCS with two DL networks, U-Net and Swinunter. The performance of U-Net and Swinunetr in our datasets were  0.5353  and 0.5603, respectively. MMCS outperformed U-Net and Swinunetr with a performance of 0.8239. 


\section{Conclusion}
\label{sec:formatting}
In this paper, we proposed three algorithms, nucleus-assisted global recognition, self-adaptive diameter filter, and temporal-ensembling models, to enhance the Cellpose-based pre-trained cytomembrane segmentation model. These algorithms enable the model to learn from both nucleus and cell morphology and facilitate efficient training with less annotated data. Our combined approach enables the model to recognize single-cell borders in multi-modality high-resolution microscopy images, resulting in an efficient cell segmentation method, MMCS.
Empirical experiment results verify the superiority that MMCS is capable of coping with diversity in microscopy images and cell shapes with the unified model, and outperforms representative single-category cell segmentation tasks. We plan to scale MMCS to a large variety of cells and microscopy in further study. In addition, more studies on mixed coenocyte segmentation will be investigated. We hope that this study will provide valuable insights into the following studies of complicated weakly supervised cell segmentation and understand how to introduce domain priori knowledge into cross-cutting DL-based microscopy image-based biology and biomedical applications.


\bibliographystyle{ieeetr}
\bibliography{mmcs}

\clearpage

\appendix

\section{More Implementation Details}

 Table~\ref{tab:enviroments1} and Table~\ref{tab:enviroments2} show more details on the development environment and training hyperparameters, respectively.

\begin{table*}[ht]
  \centering
  \begin{tabular}{@{}ll@{}}
    \toprule
  System  & Specification  \\
    \midrule
    System  & Ubuntu \\
    CPU & Intel(R) Xeon(R) Gold 6346 CPU @ 3.10GHz \\ 
    CUDA version & 11.7  \\
    Python version & Python 3.9 \\
    Deep learning framework & Pytorch \\
    \bottomrule
  \end{tabular}
  \caption{Development environments and requirements.}
  \label{tab:enviroments1}
\end{table*}

\begin{table*}[ht]
  \centering
  \begin{tabular}{@{}lll@{}}
    \toprule
  Learning setups&Pretraining&Semi-training\\
    \midrule
    Batch size&16&32 \\
    Total epoch&4000&2000-4000 \\ 
    Optimizer&SGD&SGD \\
    Initial learning rate&0.1&0.1  \\
    Lr decay schedule&N/A&Halving every 100 epochs over the last 1000 epochs \\
    Loss function&MSE, BCEWithLogits& MSE, BCEWithLogits \\
    Loss weighted schedule&N/A&Corresponding factors (0.05-0.7) for unlabeled data subsets  \\
    \bottomrule
  \end{tabular}
  \caption{MMCS  training hyperparameters.}
  \label{tab:enviroments2}
\end{table*}

\section{More Visual Results}

\begin{figure*}
  \centering
  \includegraphics[width=17cm]{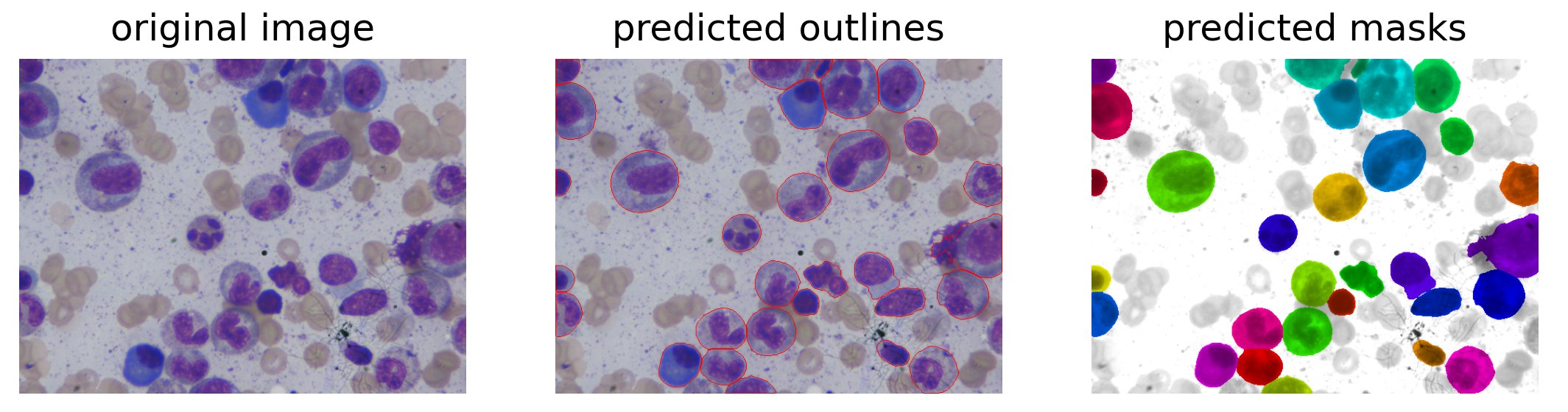}
  \includegraphics[width=17cm]{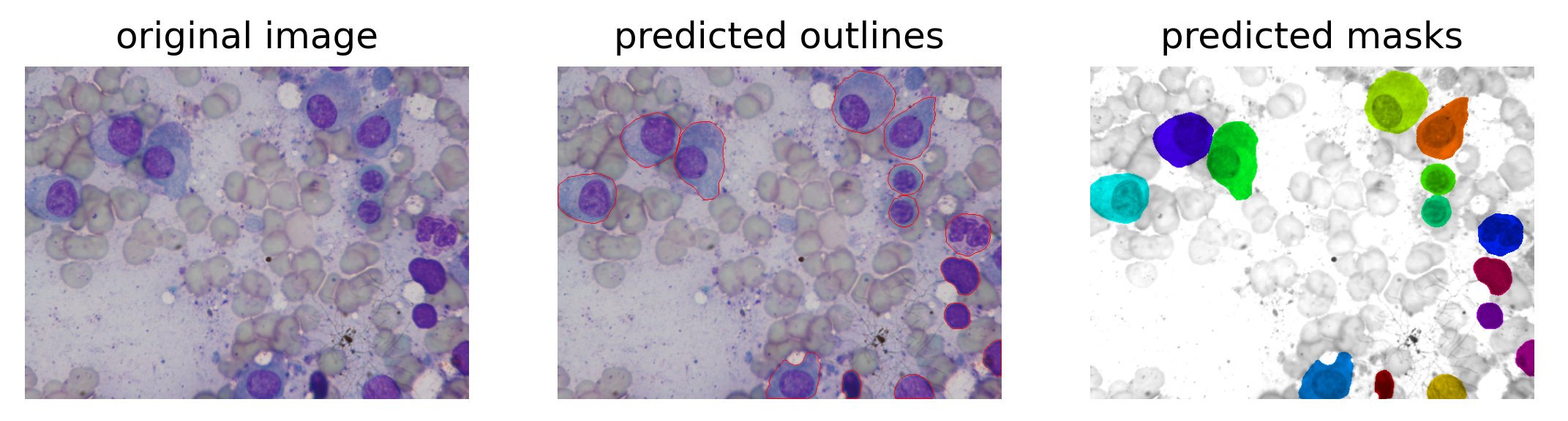}
  \includegraphics[width=17cm]{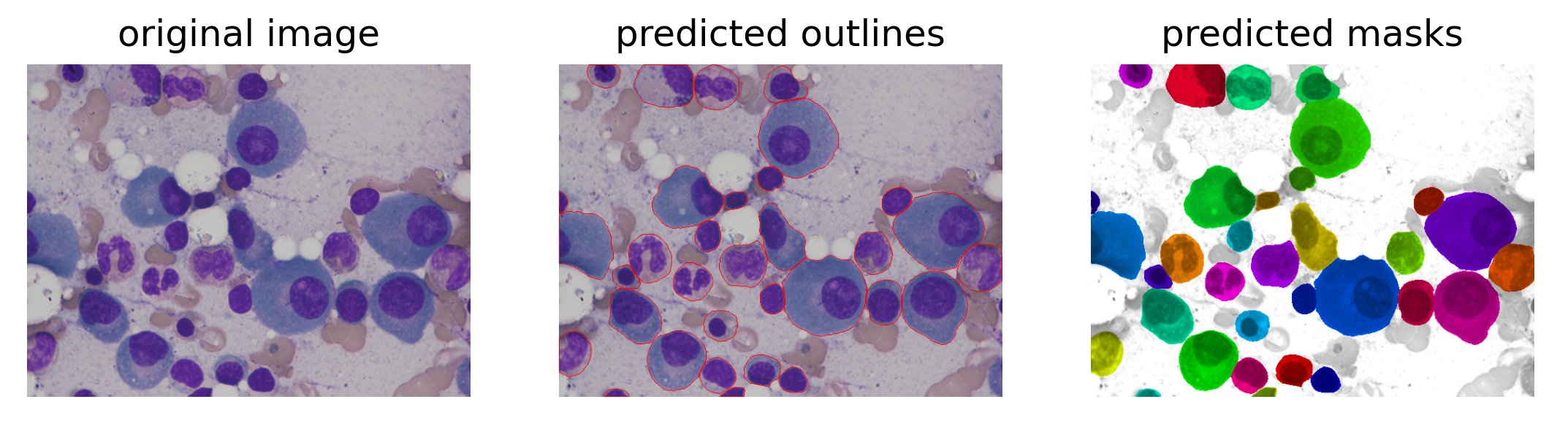}
  \includegraphics[width=17cm]{figures/SI Figure/21.jpg}
  \caption{Results on images of stained cells that originally obtained via Brightfield microscope. The predicted masks are displayed in the right panel, and the outlines in the middle panel represent the boundaries of the predicted masks. This demonstrates that the MMCS segmentation achieves human-level accuracy.}
\label{fig:V1}
\end{figure*}

\begin{figure*}
  \centering
  \includegraphics[width=17cm]{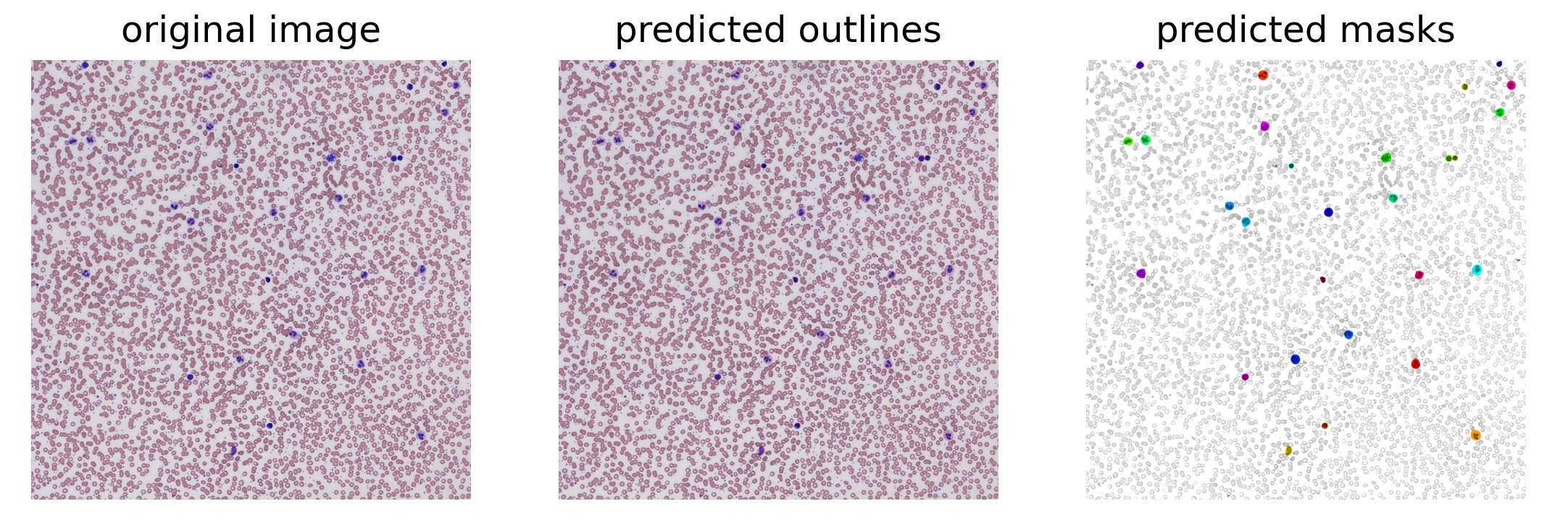}
  \includegraphics[width=17cm]{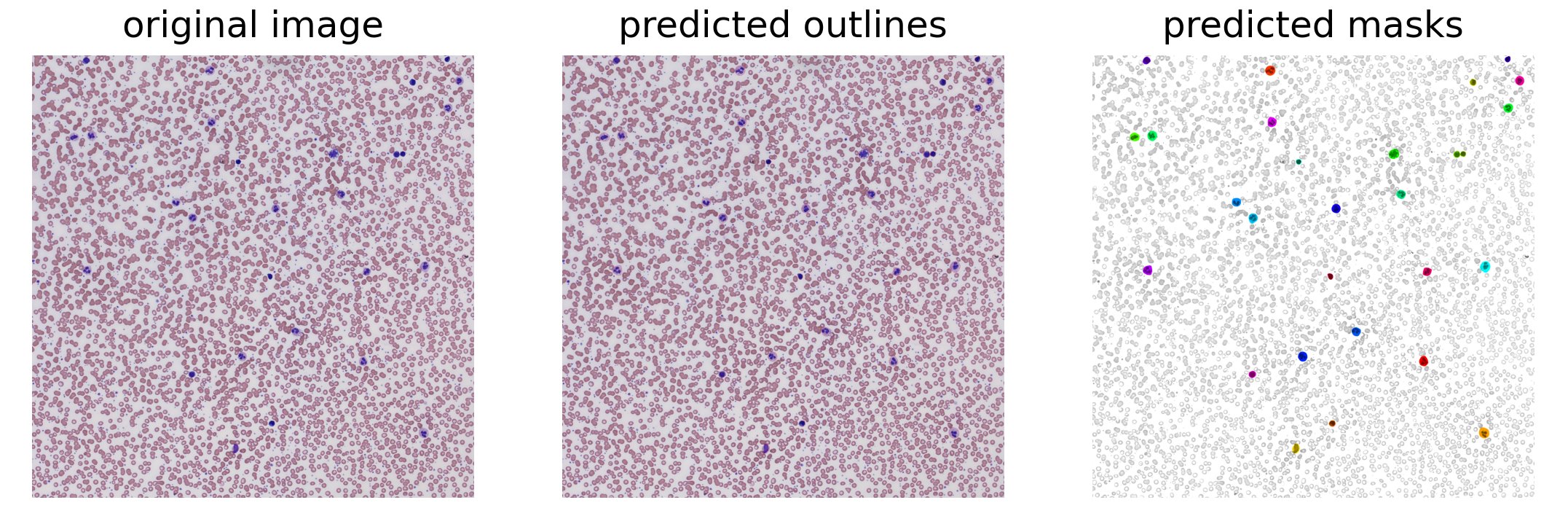}
  \includegraphics[width=17cm]{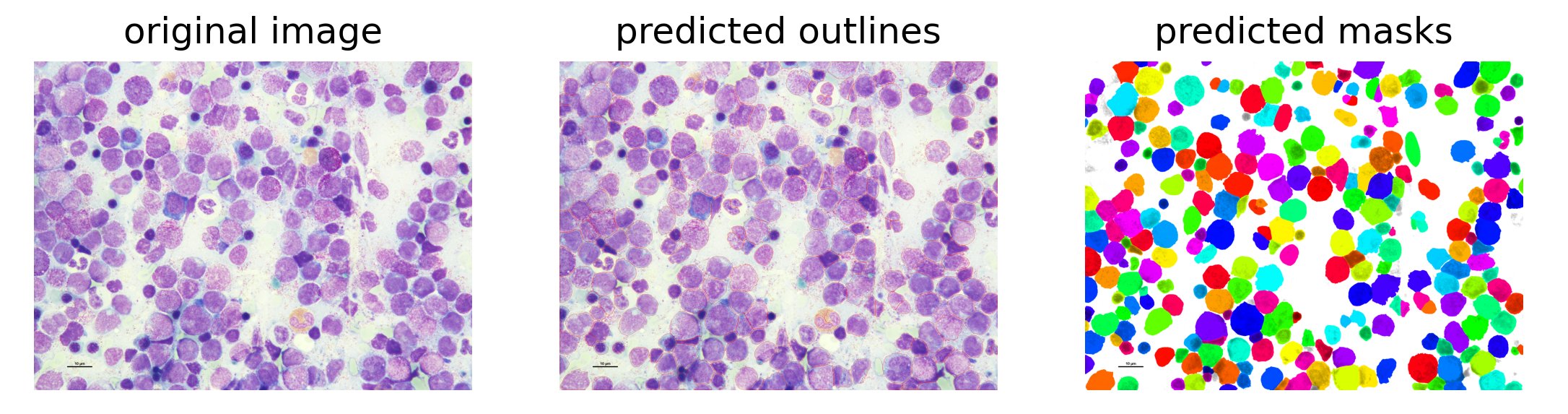}
    \includegraphics[width=17cm]{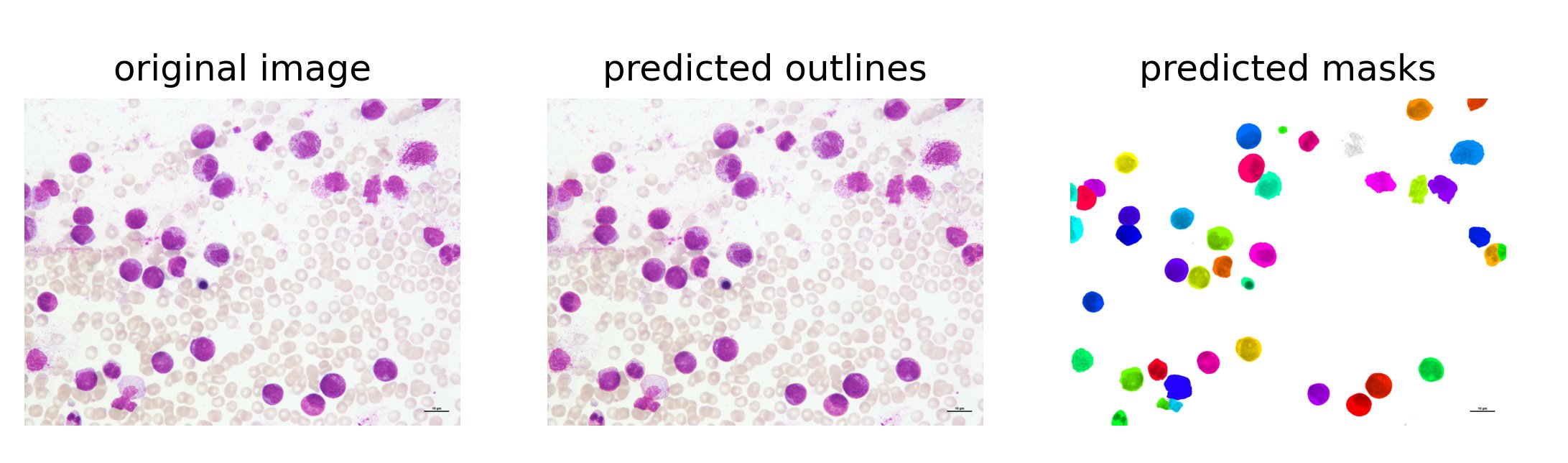}
  \caption{Results on images of stained cells that originally obtained via Brightfield microscope. MMCS performs well on stained cells even when cells are aggregated.}
\label{fig:V2}
\end{figure*}

\begin{figure*}
  \centering
  \includegraphics[width=17cm]{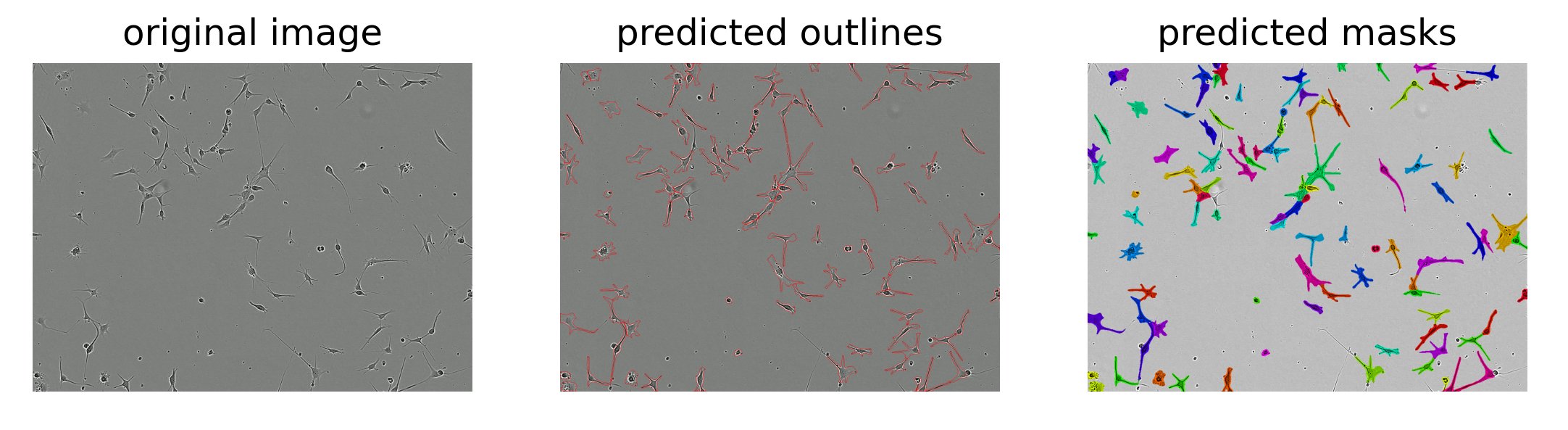}
  \includegraphics[width=17cm]{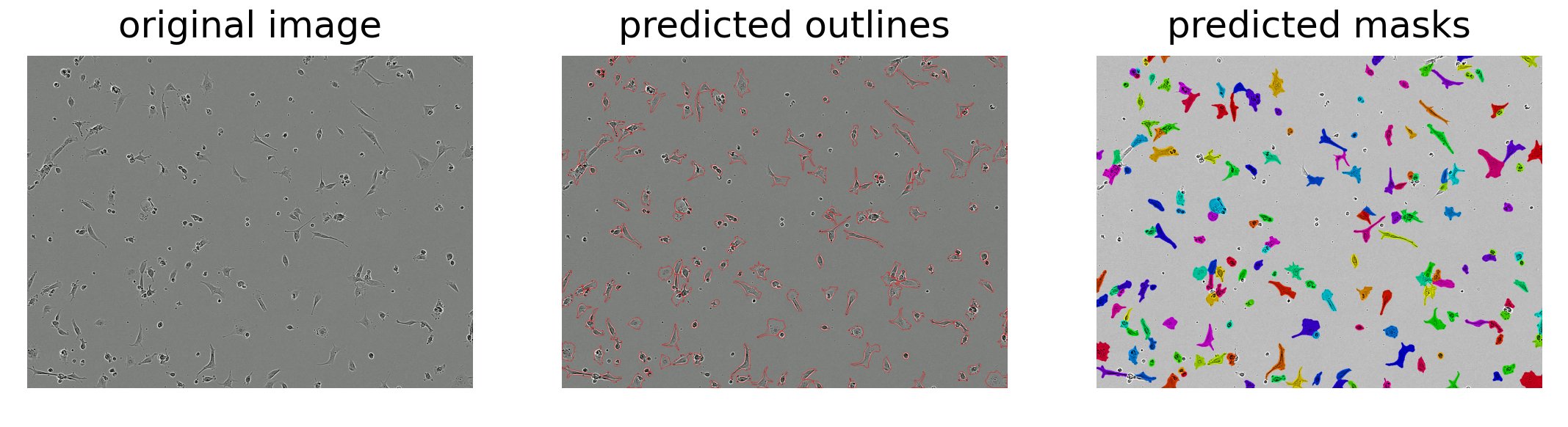}
  \includegraphics[width=17cm]{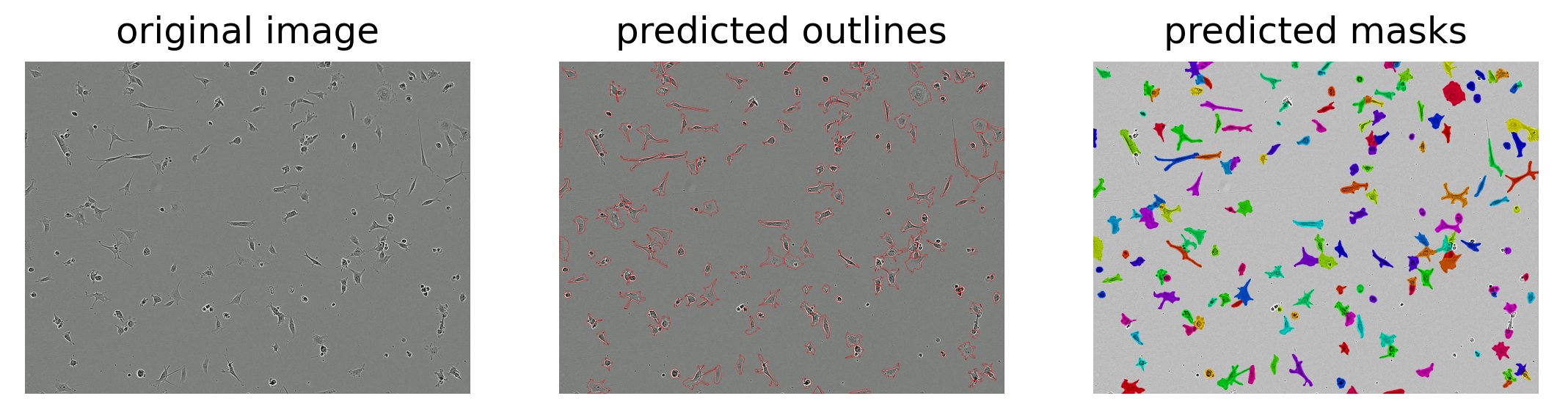}
  \includegraphics[width=17cm]{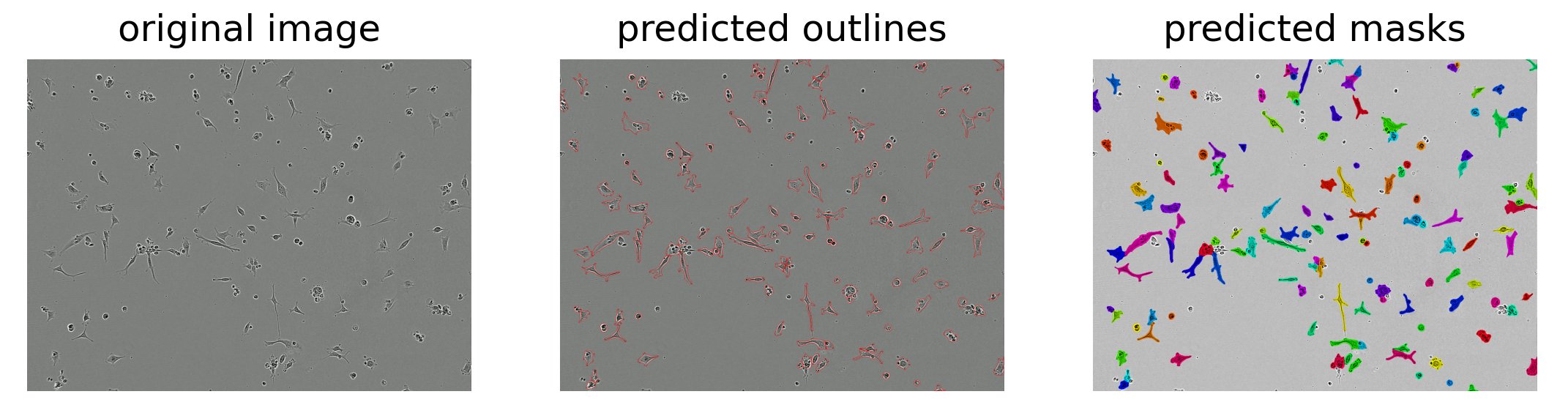}
  \caption{Results on images of cells with elongated or branched morphology that originally obtained via Phase-contrast microscope. MMCS has a good performance on cells with elongated or branched morphology.}
\label{fig:V3}
\end{figure*}

\begin{figure*}
  \centering
  \includegraphics[width=16cm]{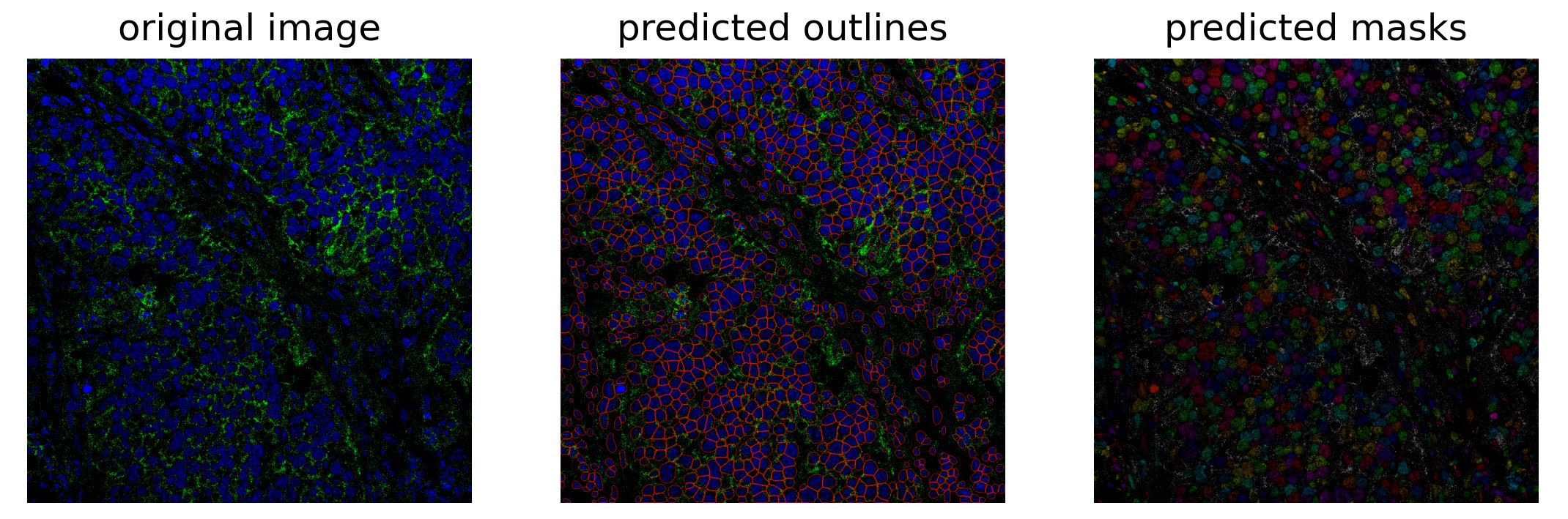}
  \includegraphics[width=16cm]{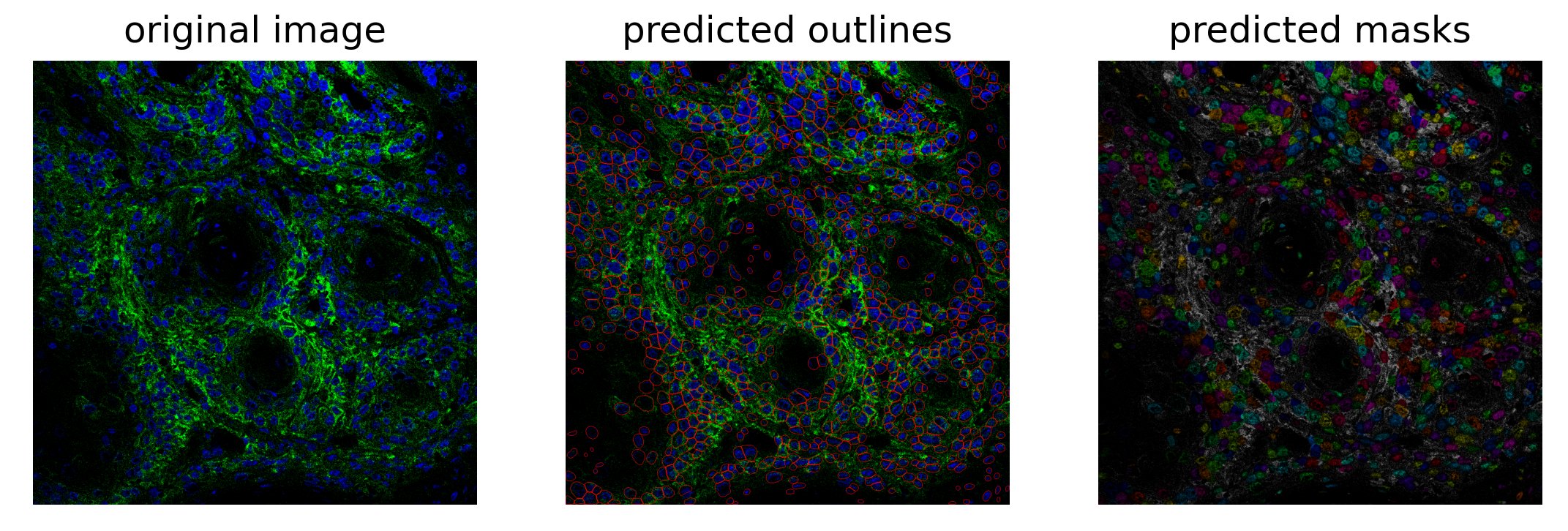}
  \includegraphics[width=16cm]{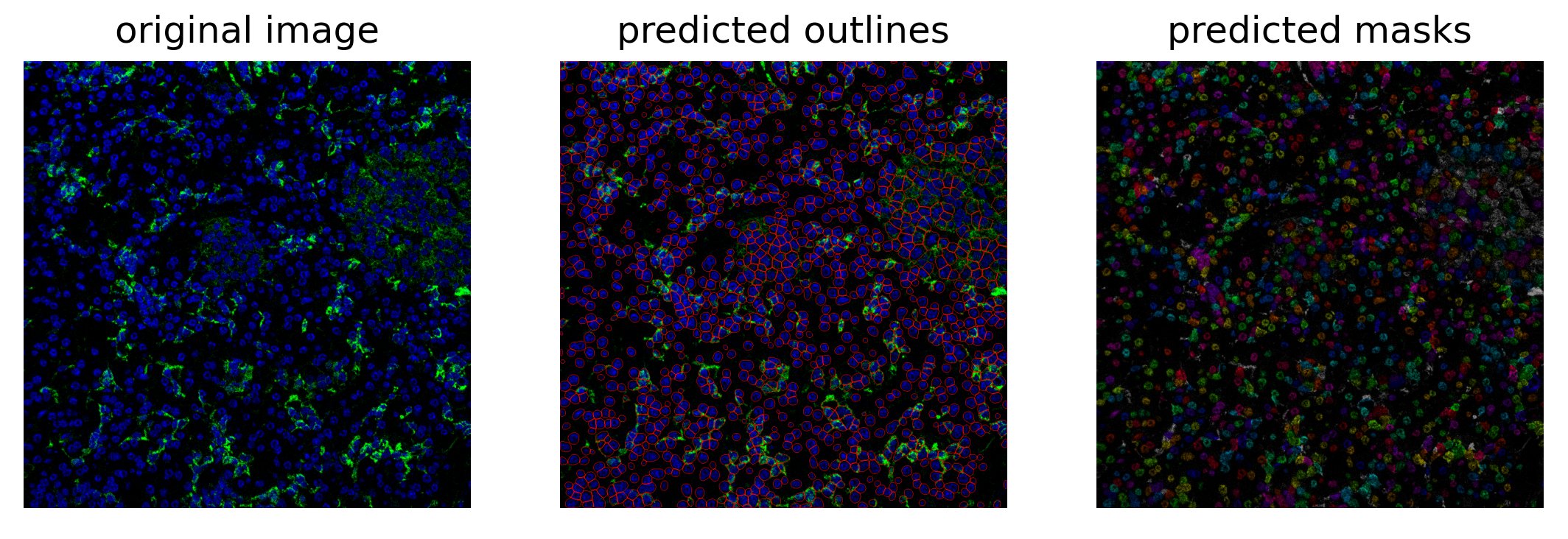}
  \includegraphics[width=16cm]{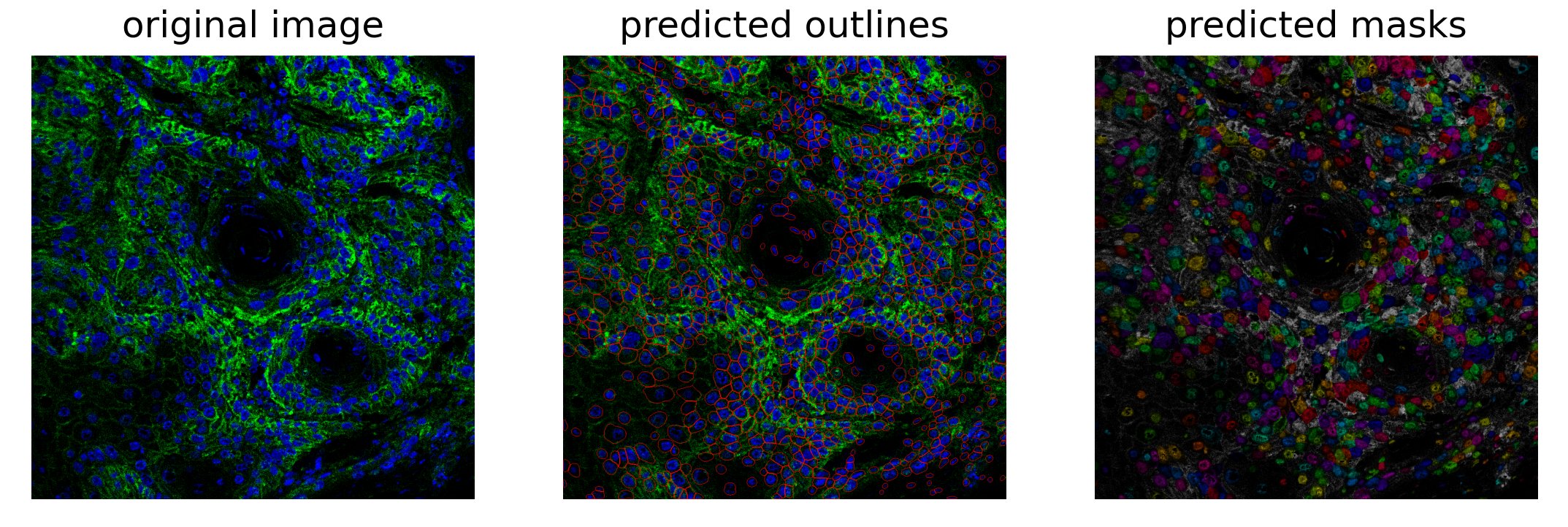}
  \caption{Results on Fluorescent images. MMCS has a good performance on fluorescent cells from tissue slices even when the cells are not separated.}
\label{fig:V4}
\end{figure*}

\begin{figure*}
  \centering
  \begin{subfigure}{0.3\linewidth}
      \centering
      \includegraphics[height=\linewidth, width=\linewidth]{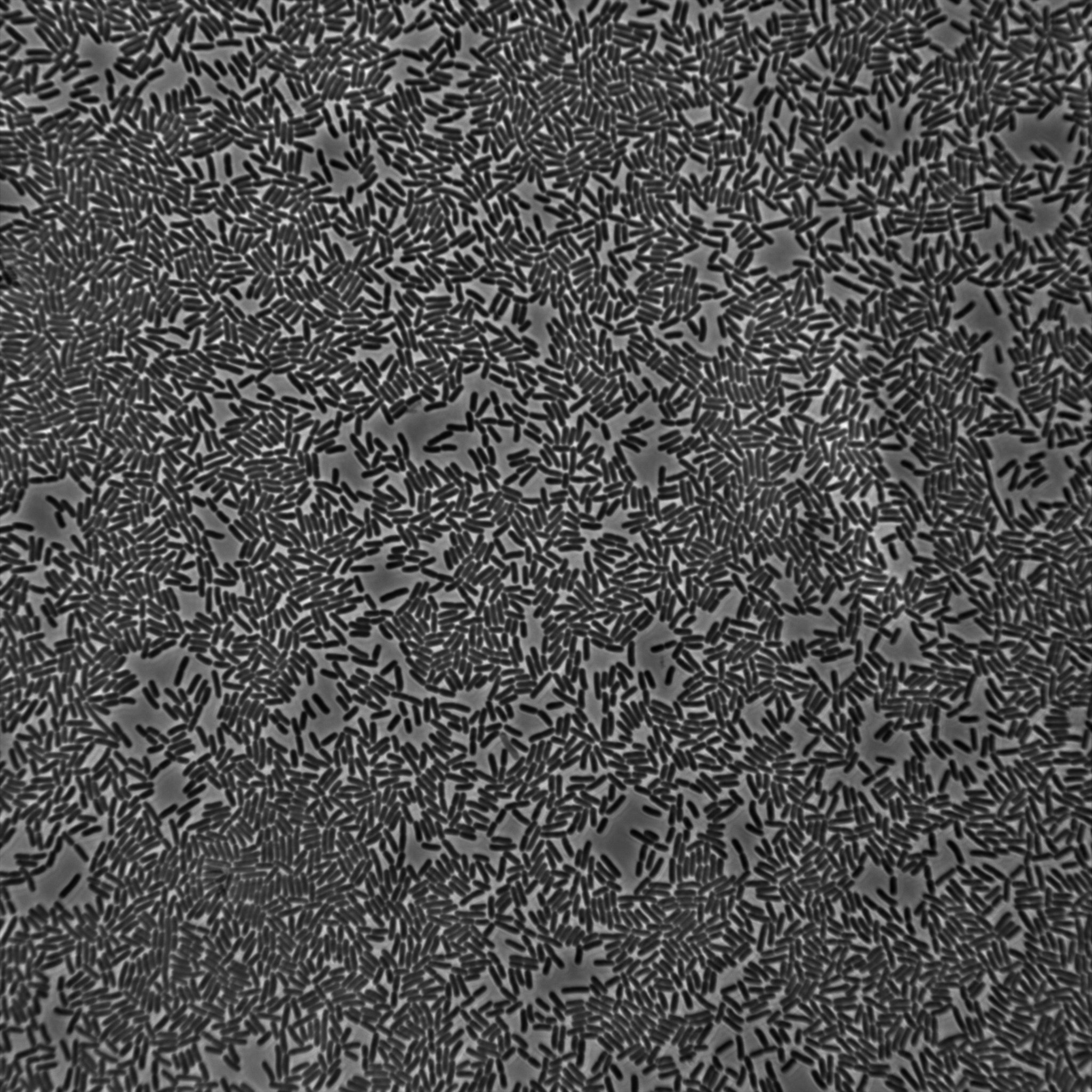}
      \includegraphics[height=\linewidth, width=\linewidth]{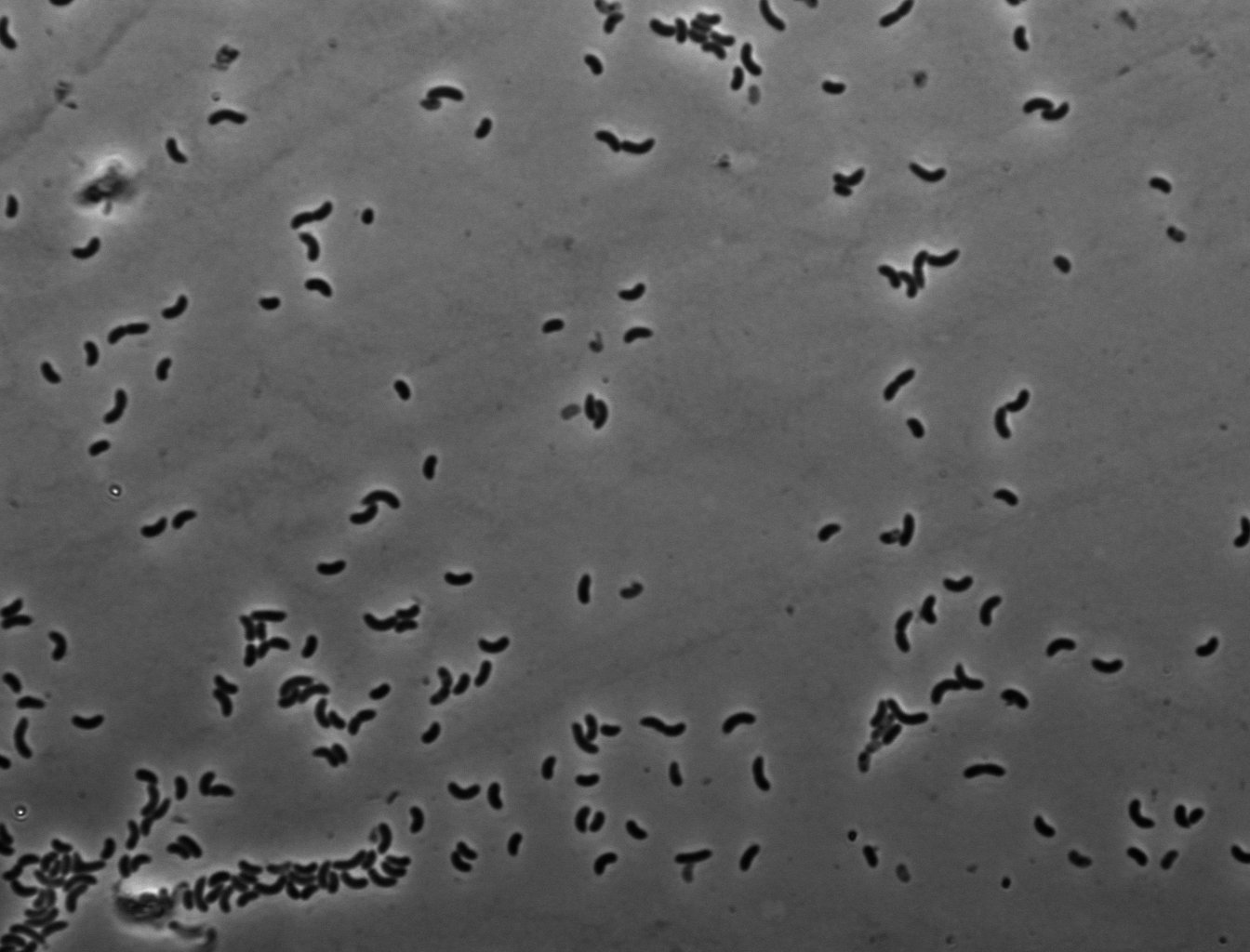}
      \includegraphics[height=\linewidth, width=\linewidth]{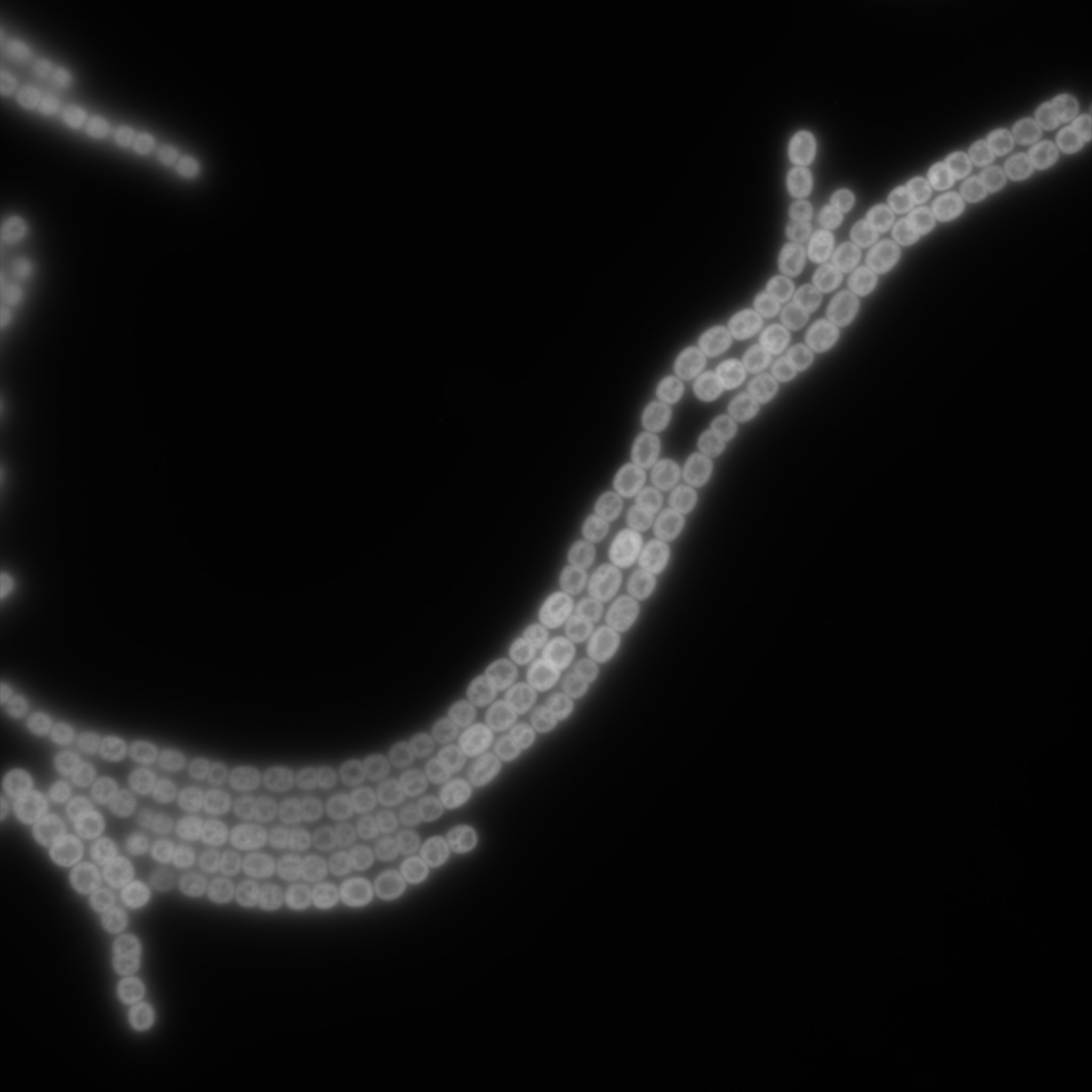}
      \subcaption{Original images}
  \end{subfigure}
  \hspace{1mm}
  \begin{subfigure}{0.3\linewidth}
      \centering
      \includegraphics[height=\linewidth, width=\linewidth]{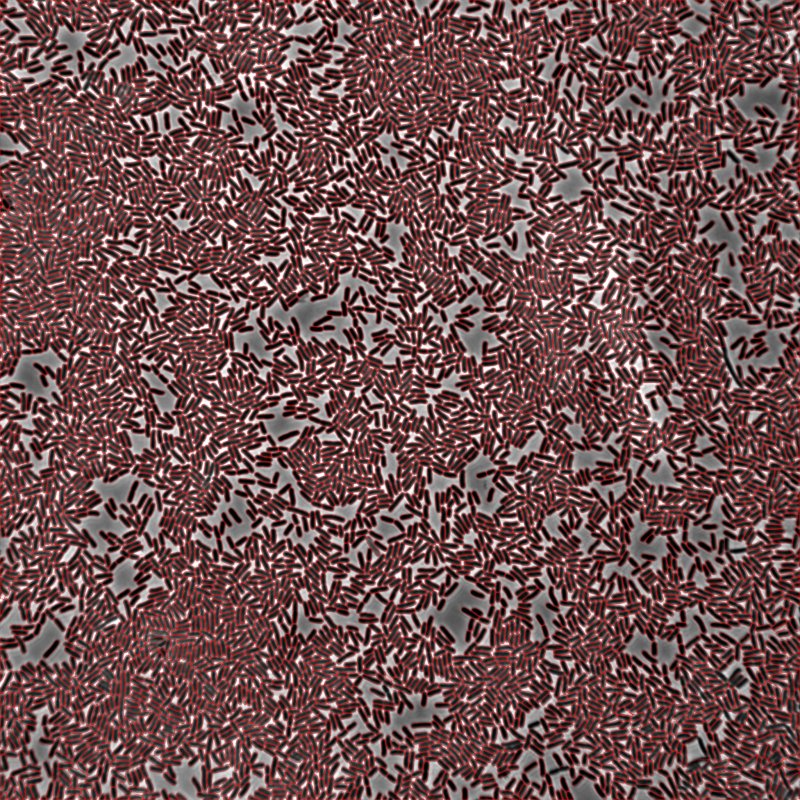}
      \includegraphics[height=\linewidth, width=\linewidth]{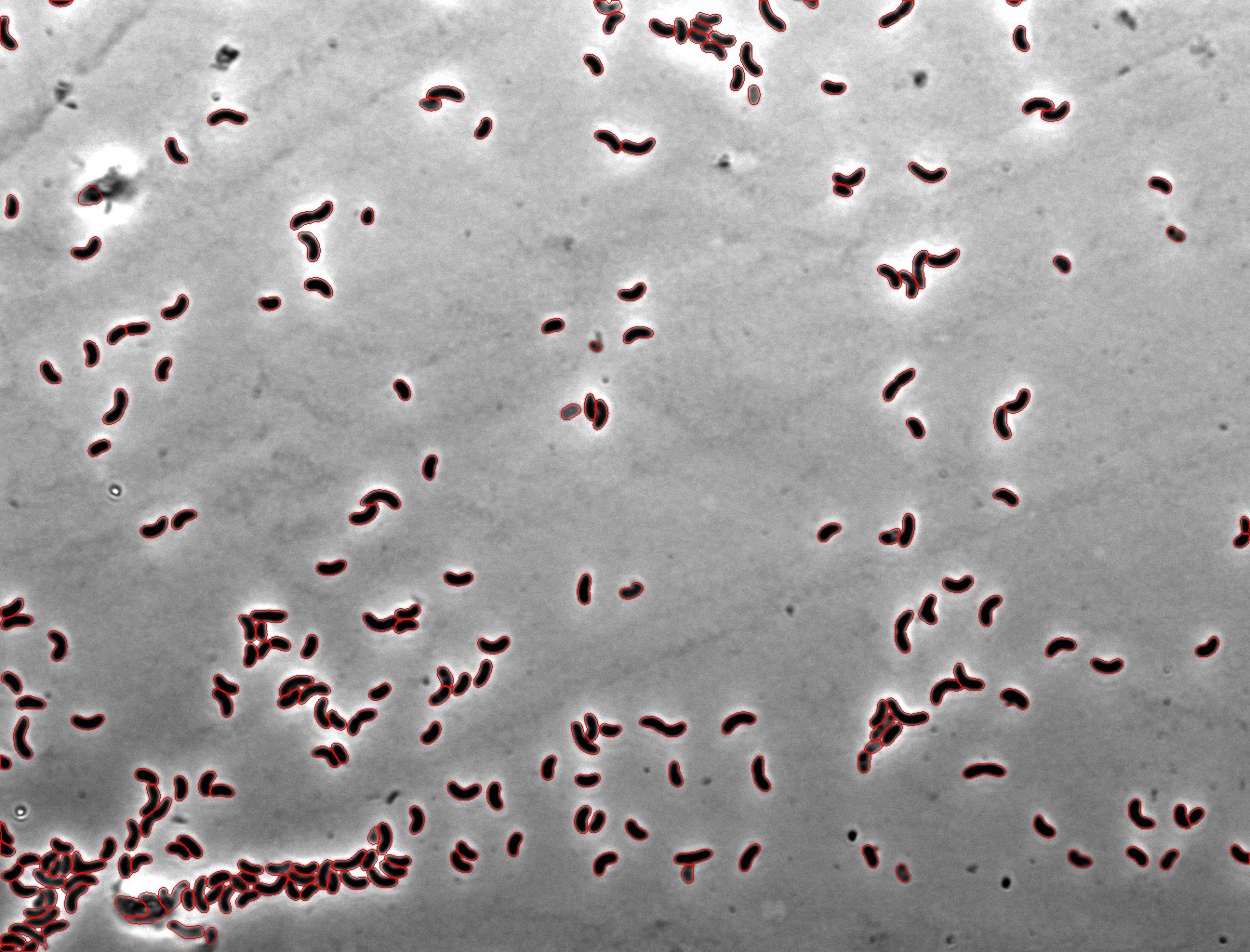}
       \includegraphics[height=\linewidth, width=\linewidth]{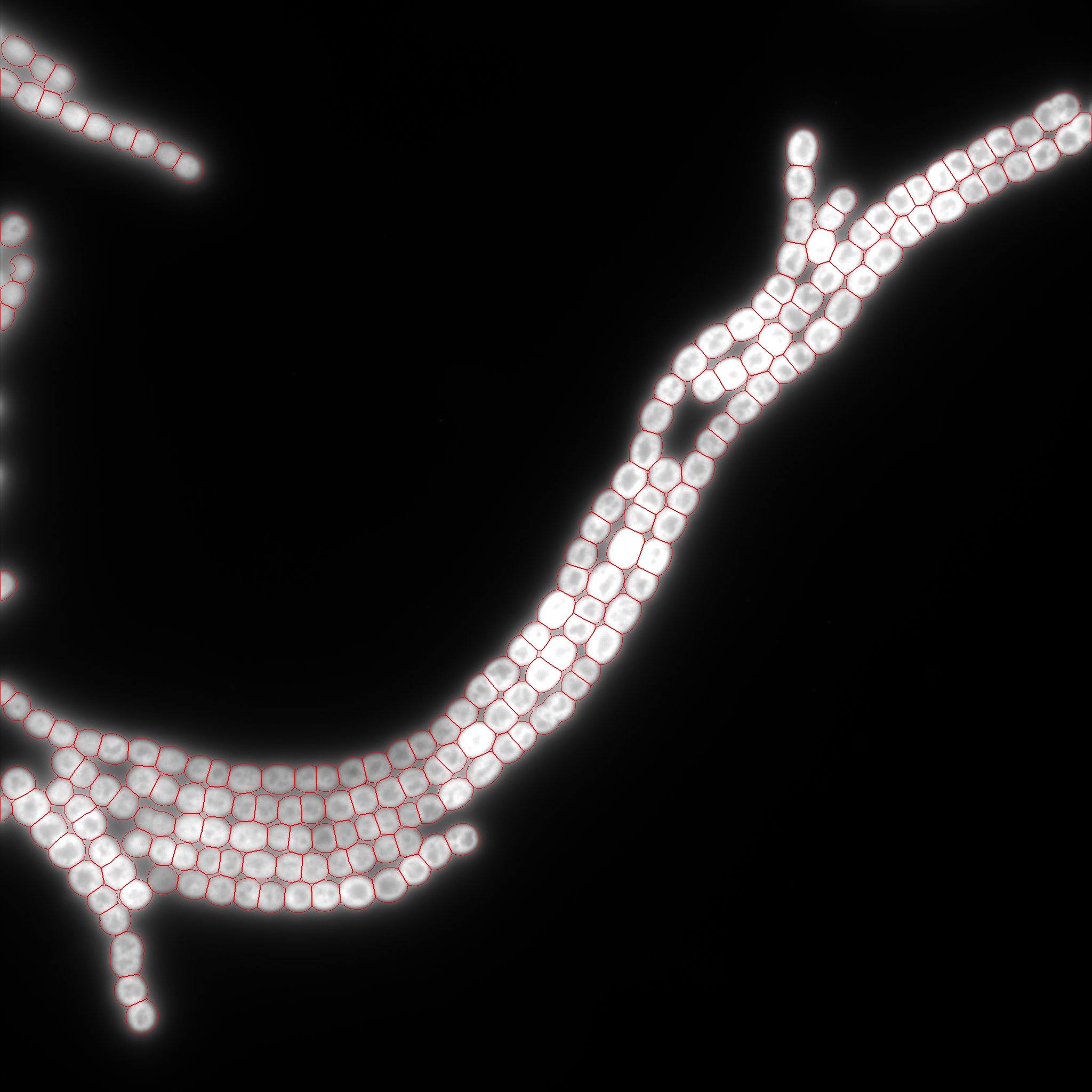}
      \subcaption{Predicted outlines}
  \end{subfigure}
  \caption{Results on images of bacterial cells that originally obtained via Differential interference contrast microscope. MMCS performs well on bacterial cells.}
\label{fig:V5}
\end{figure*}

\begin{figure*}
  \centering
  \begin{subfigure}{0.3\linewidth}
      \centering
      \includegraphics[height=\linewidth, width=\linewidth]{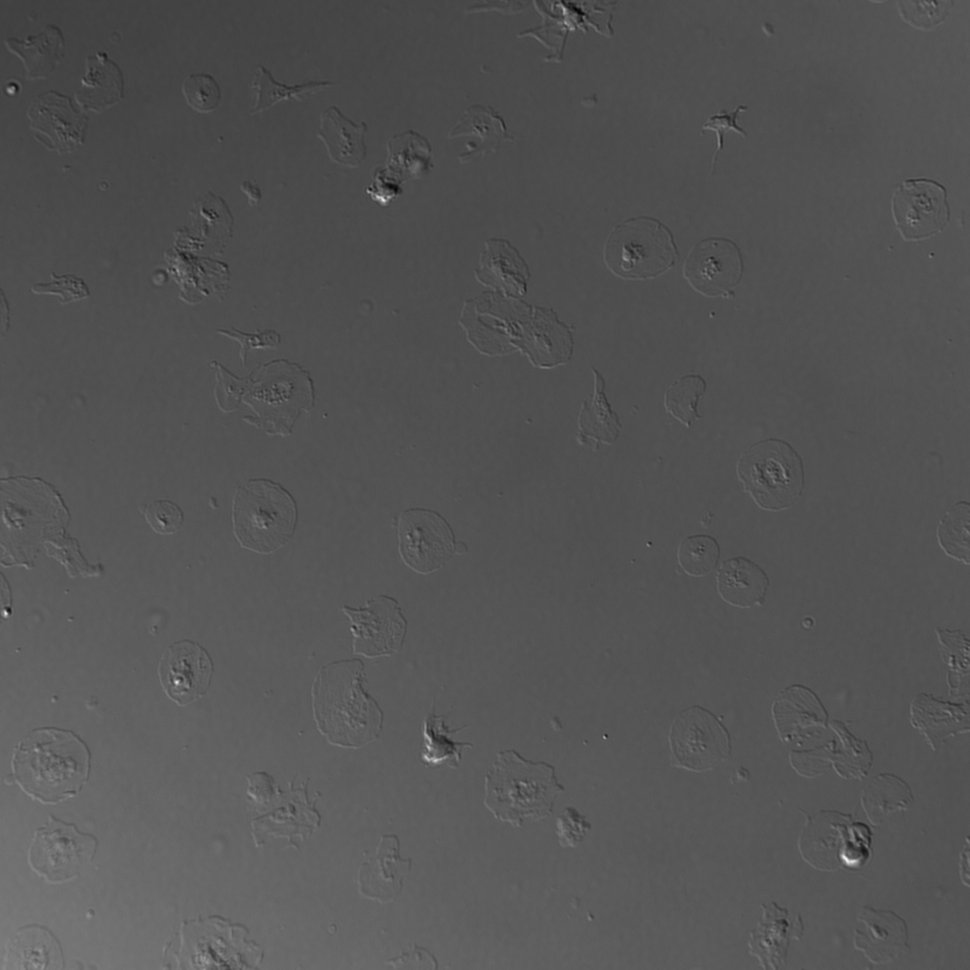}
      \subcaption{Original images}
  \end{subfigure}
  \hspace{1mm}
  \begin{subfigure}{0.3\linewidth}
      \centering
      \includegraphics[height=\linewidth, width=\linewidth]{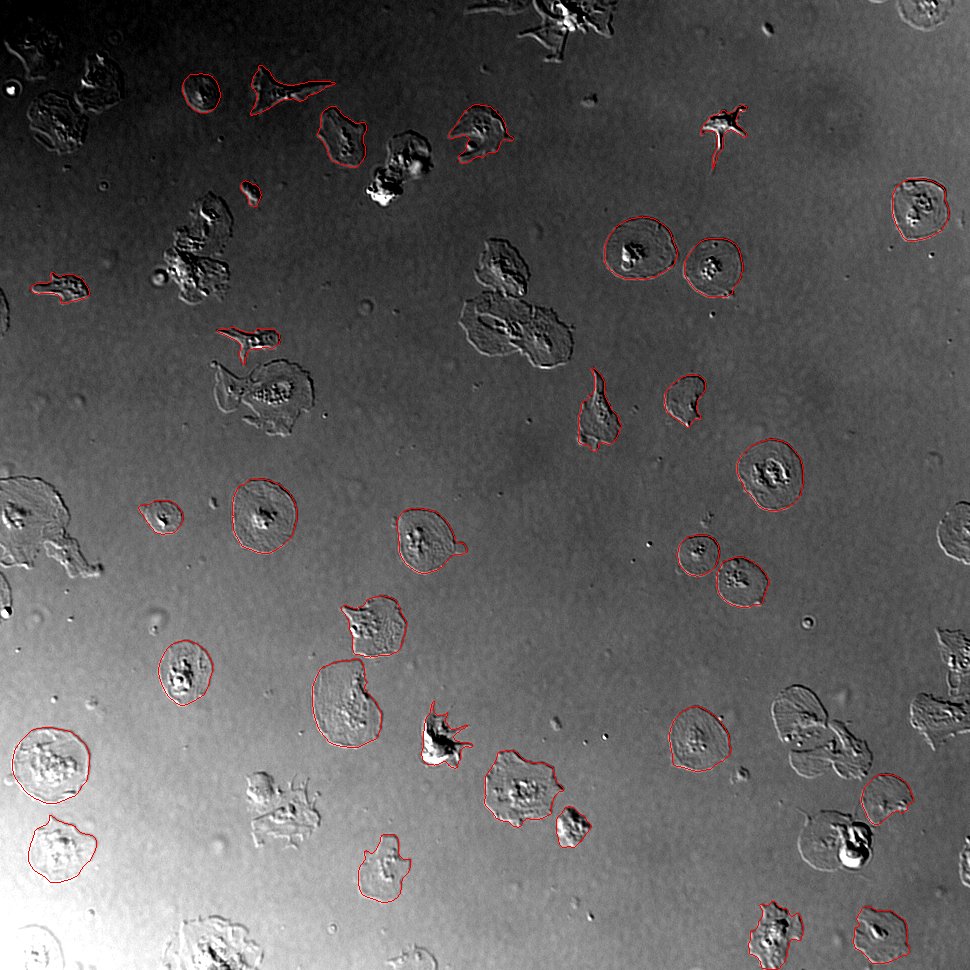}
      \subcaption{Predicted outlines}
  \end{subfigure}
  \caption{Poor segmentation results on images of live cells that originally obtained via Differential interference contrast microscope.}
\label{fig:V6}
\end{figure*}

In this section, we present more visualized experiment results, shown in Fig. \ref{fig:V1} - \ref{fig:V5}. From the visual results, it can been seen that our MMCS model performs well not only on Brightfield microscope images of stained cells mixed with weak-stained tissues, but also exhibits excellent performance on Phase-contrast microscope images of cells with elongated or branched morphology, fluorescent images, and Differential interference contrast microscope images of bacterial cells. However,  in some cell types like live cells in Differential interference contrast microscope images, MMCS may miss some cells,  shown in Fig. \ref{fig:V6}. The failures may be due to the lack of training data that belong to this type.

\end{document}